\documentclass[draft,jgrga]{AGUTeX}

\usepackage{graphicx}
\usepackage{bm}
\usepackage{color}
\usepackage{amsmath}

  \setkeys{Gin}{draft=false}
\authorrunninghead{EHAB HASSAN ET AL.}

\titlerunninghead{EQUATORIAL ELECTROJET TURBULENCE: ENERGY CASCADES}

\begin{document}

%
%

\title{Multiscale Equatorial Electrojet Turbulence: Energy Conservation, Coupling, and Cascades in a  Baseline 2-D Fluid Model}
%
%

\authors{Ehab Hassan,\altaffilmark{1,4}
 D.R. Hatch, \altaffilmark{2}
 P.J. Morrison, \altaffilmark{2}
 W. Horton\altaffilmark{2,3}
 }
\altaffiltext{1}{ICES, University of Texas at Austin, Texas, US.}
\altaffiltext{2}{IFS, University of Texas at Austin, Texas, US.}
\altaffiltext{3}{ARL, University of Texas at Austin, Texas, US.}
\altaffiltext{4}{Physics, Ain Shams University, EG.}

%
%


\begin{abstract}
Progress in understanding the coupling between  plasma instabilities in the equatorial electrojet based on a unified fluid model is reported. A deeper understanding of the linear and nonlinear evolution and the coupling of the gradient-drift and Farley-Buneman instabilities is  achieved by studying the effect of different combinations of the density-gradient scale-lengths ($L_n$) and cross-field ($\bm{{E}\times{B}}$) drifts on the plasma turbulence.  Mechanisms and channels of energy transfer are illucidated for  these multiscale instabilities.  Energy  for the unified model is  examined, including the injected, conservative redistribution (between fields and scales), and ultimate dissipation. Various physical mechanisms involved in the energetics are categorized as sources, sinks, nonlinear transfer, and coupling to show that the system satisfies the fundamental law of energy conservation. The physics of the nonlinear transfer terms is studied to identify their roles in producing energy cascades -- the transference of  energy from the dominant unstable wavenumbers and driving irregularities of different scale-lengths.  The theory of two-step energy cascading to generate the 3-meter plasma irregularities in the equatorial electrojet is verified. In addition, the nonlinearity of  the system allows the  possibility for a reverse energy cascading,  responsible for generating large-scale plasma structures at the top of the electrojet.

\end{abstract}

%
%

%

\begin{article}
%
%
\section{Introduction}\label{section:Introduction}
The mixing of turbulent structures of different scale-lengths is important for understanding plasma turbulence driven by instabilities in the ionosphere. In the equatorial electrojet, Farley-Buneman and gradient-drift instabilities are driven by a large eastward electrojet current and an  $\bm{{E}\times{B}}$ drift due to an electron density-gradient, respectively [Farley 2009]. Echoes of Farley-Buneman instability are called ``Type-I" because they were the first observed using the coherent backscattering radar at Jicamarca by Bowles 
{\it et al.}~[1963]. However,  echoes for gradient-drift instability are observed only during the absence of the Type-I instability (when the electron drift speed is less than the local ion-acoustic speed, $C_s$) and are called ``Type-II" [Balsley 1969, Fejer {\it et al.}~1975]. A unified fluid model that describes both types of plasma instabilities in the equatorial electrojet under various solar conditions was discussed by Hassan \textit{et al.}~[2015].  Simulation results of this unified fluid model elucidated some facets of  the nature of  the coexistence and coupling between Farley-Buneman and gradient-drift instabilities in both the linear and nonlinear regimes.  The present work is a continuation of this previous study.\\
\hfill\\
\textit{In-situ} measurements taken by a sounding rocket launched from Lobos, Peru revealed  electrostatic characteristics of the E-region instabilities [Pfaff {\it et al.}~1987a]. These measurements also showed maximum variations of the spectral energy at frequencies less than 100 Hz at 90-103 km altitudes which extends to higher frequencies ($\sim$ 1000 Hz) at 103-108 km altitudes. Simultaneously, a backscattering radar measurement taken at Jicamarca, Peru showed a strong Type-I echo of 3 m wavelength which is modified by long-scale horizontally propagating Type-II waves [Kudeki {\it et al.}~1987]. In Hassan \textit{et al.}~[2015] the linear growth-rate profile in the vertical direction of the ionosphere was seen to have three distinct regions in the equatorial electrojet:   The lower region (90-103 km) being dominated by the gradient-drift instability, the higher region (108-115 km) being  dominated by Farley-Buneman instability, and the middle region (103-108) showing a strong coupling between the two types of plasma instabilities. Under varying ionospheric conditions, linear and nonlinear simulation results that reproduce many of the radar and rocket observations were described in that  paper and  light was shed on the physical mechanisms that drive these instabilities and are responsible for the coupling between them.\\
\hfill\\
Theoretically, Sudan {\it et al.}~[1973] used a two-step energy cascading mechanism to explain the generation of meter scale Type-II irregularities detected by 50-MHz incoherent scattering radar at electron drifts of order 100 m/s. The gradient-drift instability is  excited in the daytime with positive gradient and westward drift of the electrons, but when the amplitude of large-scale instability wave reaches a certain level, the energy starts to transfer from that primary long-wavelength waves to secondary short-wavelength waves. This energy transfer takes place due to the strong perturbations of the electrojet local parameters, where the horizontal density gradient becomes greater than the background vertical gradients and the magnitude of the vertically perturbed drifts increases up to the order of magnitude of the horizontal electron drift [Farely and Balsley 1973]. However, the excitation of vertical irregularities of wavelength of the order of meters or sub-meter is a pure Type-I nonlinear mechanism that takes place only when the electron drifts exceed the ion-acoustic speed [Sudan {\it et al.}~1973]. The nonlinear simulation results that are presented in this present paper  show a redistribution of energy from the large structures to the small ones in the vertical direction,  which verifies  the two-step energy cascading mechanism proposed by Sudan
\textit{et al.}~[1973].\\
\hfill\\
Nonlinear coupling between two unstable modes propagating in nearly the same direction can give rise to a forward-cascading of energy from  large-structures to the short ones. However, the nonlinear interaction between two unstable structures of short wavelength might cause an inverse-cascade of energy to generate a stable large-scale structure. The nonlinear coupling mechanisms between the linearly unstable modes were also found to be sensitive to the aspect angle of the equatorial electrojet irregularities [Kudeki and Farley, 1989 and Lu {\it et al.}, 2008]. Studying the energy cascading in our unified fluid dynamical system shows the ability of the gradient-drift instability to generate small-scale instability of the order of a meter and less in a forward energy cascading process.  There also exists, via a separate nonlinear mechanism, a smaller nonlinear transfer channel whereby the Farley-Buneman instability transfers energy from the sub-meter structures to irregularities of tens of meters scale-size through a reverse energy cascading mechanism.\\
\hfill\\
Our paper is organized as follows.  In section \ref{section:unifiedModel} we present the unified fluid model equations, including  some linear and nonlinear numerical results in sections  \ref{subsection:Linear} and \ref{subsection:Nonlinear}, respectively.  Next, in section \ref{section:EnergyConservation},  we present the energy equation for the evolving fields.   Here we separate out terms that correspond to sources, sinks, nonlinear transfer, and coupling.  Then, in section \ref{subsection:TotalEnergy},   we examine the total energy content and its temporal variations in our simulations, followed in section \ref{subsection:EnergySourcesSinks}, by an investigation of the roles of sources and sinks of energy in the system.  This is followed up in section \ref{section:EnergyCascades} where the transfer of energy, cascading,  among the unstable modes is investigated.  Finally, in  section \ref{section:SummaryConclusions} we summarize our findings.
\section{Unified Fluid Model}\label{section:unifiedModel}
As noted above, in  the previous work of Hassan \textit{et al.}~[2015] the  2-D unified fluid model that unifies the Farley-Buneman and gradient-drift instabilities was derived. There it was shown that this  model has linear results that are comparable to the kinetic treatment of the Farley-Buneman plasma instability by Schmidt and Gary [1973], and simulation results also showed good agreement with  radar and sounding rocket observations. The unified fluid model is comprised of the following set of nonlinear partial differential equations that control  the plasma dynamics and energy coupling between the unstable modes:
\begin{eqnarray}\label{eq:Uni3fIonContinuity}
\hspace{1.5 cm} \partial_tn&=&\bm{\nabla}\cdot\left(n\bm{\nabla}\chi\right)
\\
\label{eq:Uni3fIonMomentum}
\partial_t\chi&=&\upsilon_{t_i}^2\ln(n)+\frac{\Omega_{ci}}{B}\phi-\nu_{in}\chi+\frac{1}{2}|{\bm\nabla}\chi|^2+\frac{4}{3}\frac{\upsilon_{t_i}^2}{\nu_{in}}\nabla^2\chi
\\
\partial_t\nabla^2\phi&=&\frac{T_e\nu_{en}}{e}\nabla^2\ln(n)-\nu_{en}\nabla^2\phi-B\Omega_{ce}\nabla^2\chi\nonumber\\
                      &&\qquad -\ \Omega_{ce}\left[\phi,\ln(n)\right]-\frac{1}{B}\left[\phi,\nabla^2\phi\right]\nonumber\\
                      &&\qquad +\ \frac{T_e\nu_{en}}{e}{\bm\nabla}\ln(n)\cdot{\bm\nabla}\ln(n)-\nu_{en}{\bm\nabla}\ln(n)\cdot{\bm\nabla}\phi\nonumber\\
                      &&\qquad -\ B\Omega_{ce}{\bm\nabla}\ln(n)\cdot{\bm\nabla}\chi,
 \label{eq:Uni3fQuasineutrality}                  
\end{eqnarray}
where $\upsilon_{t_i}$ is the ion thermal speed, $\nu_{in}$ and $\nu_{en}$ are, respectively,  the ion and electron collision frequencies with the background neutrals, $\Omega_{ci}$ and $\Omega_{ce}$ are the ion and electron gyration frequencies, $T_e$ is the isothermal temperature of ions and electrons, $B$ is the background magnetic field, and $\left[f,g\right]$ is the Poisson  bracket (Jacobian) which is defined as $\left[f,g\right]=\partial_x{f}\partial_y{g}-\partial_x{g}\partial_y{f}$.\\

The three dynamical variables of the unified fluid model are $(n, \chi,\phi)$.  Specific details regarding the justification for this model are given in Hassan \textit{et al.}~[2015]; here we note two key ingredients. The variable $n$ is the plasma density of either species,  since quasineutrality  has been assumed.  Because of the large ratio between the ion collision frequency with the neutral particles and the ion gyration frequency [Kelley 2009]  the following approximation for the ion fluid velocity was assumed:
\begin{center}
$\bm{\upsilon}_i$ = $-\bm{\nabla}\chi$.  
\end{center}
Because of the small temporal and spatial variations in the background magnetic field during the solar quiet times [Farley 2009] and because of the electrostatic nature of the plasma waves in the equatorial electrojet,   the ${\bf{E}}$-field was assumed to be  of the form 
\begin{center}
${\bf{E}}=-{\bm\nabla}\phi$. 
\end{center}

The linear and nonlinear simulation results of this model reproduce many features that are found in the radar and rocket observations as we will show below.

\subsection{Linear Results}\label{subsection:Linear}
\begin{figure}
 \centering
 \noindent\includegraphics[width=1.0\textwidth]{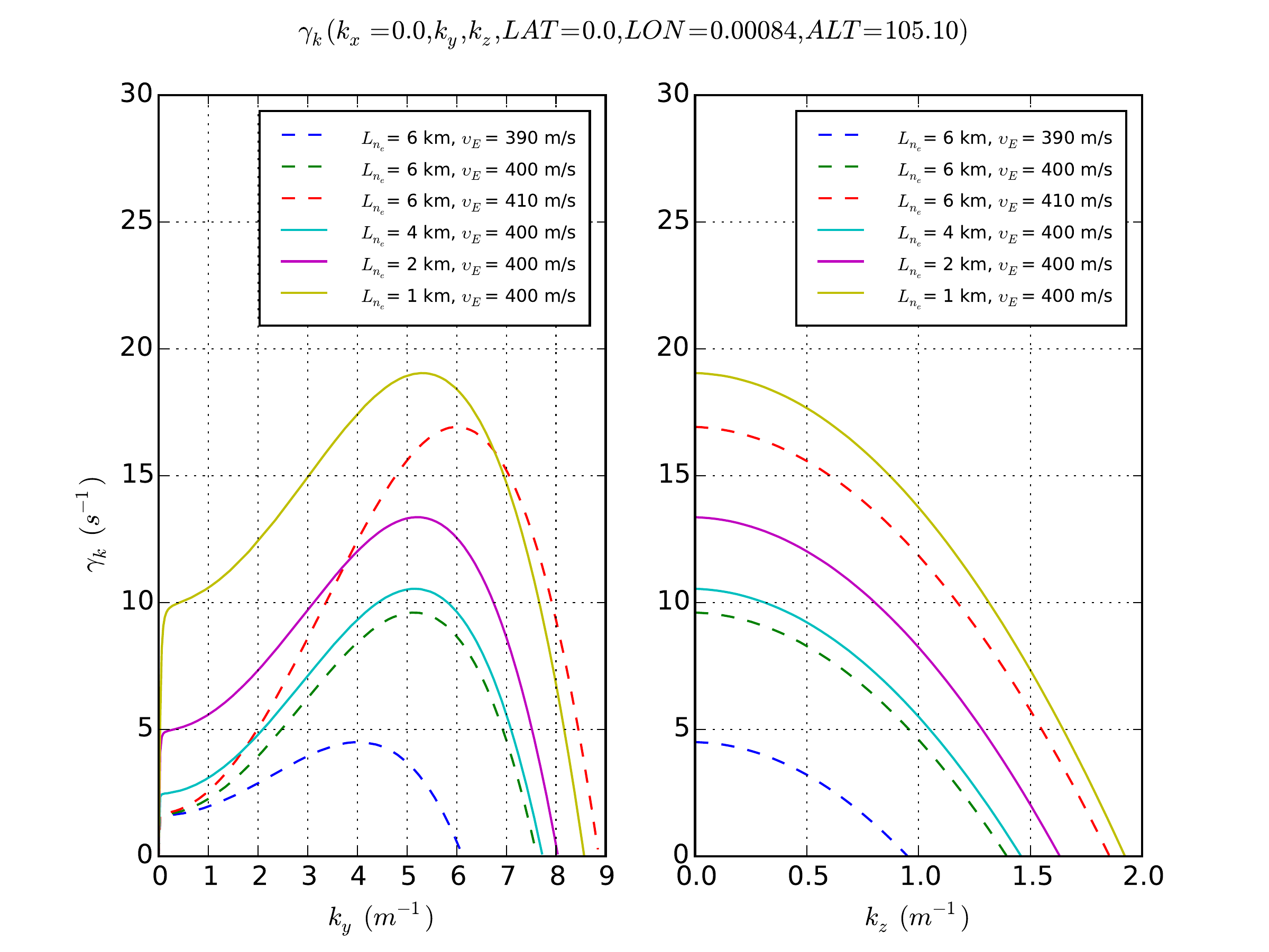}
 \noindent\caption{\small The dependence of the growth-rates at 105 km altitude on the horizontal ($k_y$) (left panel) and vertical ($k_z$) (right panel) wavenumbers are shown for different values of the electron density scale-lengths $L_n$ (top) and cross-field ($\bm{{E}\times{B}}$) drifts.}
 \label{fig:Growthrates}
\end{figure}

The dynamical system described by  equations (\ref{eq:Uni3fIonContinuity}), (\ref{eq:Uni3fIonMomentum}), and (\ref{eq:Uni3fQuasineutrality}) can be linearized by separating out the stationary background from the fluctuating part as: {$\tilde{n}=n_o+\delta\tilde{n}$, $\tilde{\phi}=\phi_o+\delta\tilde{\phi}$,  and  $\tilde{\chi}=\chi_o +\delta\tilde{\chi}$}, where $\nabla{\chi_o}=0$. Then we can solve the linearized equations numerically to find the eigenvalues and eigenvectors for our linear system, giving as usual  the growth-rate and frequency (and by extension, phase velocity) as the imaginary and real parts of the eigenvalues, respectively.\\
\hfill\\
In Figure \ref{fig:Growthrates}, the growth-rate profiles are shown as a function of $k_y$ in the left panel and $k_z$ in the right panel. Choosing $k_{min} = \pi/16$ $m^{-1}$ for both the horizontal and vertical directions gives rise to a single peak in $\gamma(k_z)$ and two peaks in $\gamma(k_y)$, with one located at small values of $k_y$. In the right panel, it is seen that there is no minimum or threshold wavenumber  for exciting  unstable modes and the  peak in $\gamma(k_z)$ is found at $k_z = 0$. This means that unstable modes are always present in the vertical direction as a result of the electron $\bm{{E}\times{B}}$ drift  $\upsilon_E$ and a positive density-gradient scale-length $L_n$. However, examination of the left panel reveals two peaks in the $k_y$-dependent growth-rate.  Observe  a short peak  at  small $k_y$ near  $k_y = \pi/16$ $m^{-1}$,  which manifests itself by generating plasma waves of long-wavelength. The  clear maximum of $\gamma(k_y)$ at larger $k_y$ can be associated with the linear excitation of short-scale (sub-meter) plasma structures.\\
\hfill\\
The electron density-gradient scale-length and $\bm{{E}\times{B}}$ drifts are the salient parameters in driving the plasma instabilities at small and large wavenumbers. The solid-lines in Figure \ref{fig:Growthrates} show the effect of different $L_n$ (at a constant drift speed, $\upsilon$ = 400 m/s) in driving the instabilities at small-wavenumbers for the  $\gamma(k_y)$ and $\gamma(k_z)$ profiles. The short density-gradient scale-length gives rise to a higher peak at the small-wavenumbers for the growth-rate profiles in the horizontal and vertical directions. In addition, this short $L_n$ affects the height of the $\gamma(k_y)$ peak at large-wavenumbers. Therefore, the density-gradient scale-length plays a very important role in driving equatorial electrojet instabilities for structures of different scale-lengths.\\
\hfill\\
On the other hand, the electron's $\bm{{E}\times{B}}$ drift  only affects the height of the growth-rate profile in the horizontal direction $\gamma(k_y)$ at the large-wavenumbers (small structures), where the drift velocity is observed to increase growth rates. In some other cases (not shown here) we have examined the growth rates when $\bm{{E}\times{B}}$ drifts are smaller than the local ion-acoustic speed. In these cases, the peak at large-wavenumbers in the $\gamma(k_y)$ profile disappears, but the one at small-wavenumbers does not. Therefore, the instability at the large-wavenumbers is due to the two-stream instability that is excited when the relative speed between the electrons and ions exceeds the ion-acoustic speed. Also, there is no minimum threshold for the cross-field drift to excite the gradient-drift instability at the small wavenumbers, although  it has to be greater than zero.\\
\hfill\\
Another representation of the variation of the growth-rate with the density-gradient scale-length $L_n$ and the cross-field drift $\upsilon_E$ is shown in Figure \ref{fig:Growth2D}. Note the positive growth-rate for all values of $L_n$ and $\upsilon_E$ which proves the presence of at least one type of instability (i.e.\  Type-II) in the equatorial electrojet region even in the absence of Type-I instability when the cross-field drift speed is less than the ion-acoustic speed. In the region below the ion-acoustic speed the growth-rate of Type-II instability is influenced by variations in the density-gradient scale-length $L_n$ and the cross-field drift speed $\upsilon_E$. Below $L_n$ = 1.5 km, the scale-length dominates and the change in the growth-rate due to the cross-field drift is negligible. In contrast, above $L_n$ = 4.0 km, the variation in growth-rate due to the change in the density-gradient scale-length is very small compared to the effect of the variation of the cross-field drift on the growth-rate. In the transition region ($L_n$ = 1.5 - 4.5 km), the variation in the growth-rate is influenced by the variations in $L_n$ and $\upsilon_E$. Thus, the plasma dynamics of Type-II instability in the equatorial electrojet depends on the background parameters in the three ranges indicated above.\\
\hfill\\
In the presence of Type-I instability, when the cross-field drift exceeds the ion-acoustic speed, the effect of $\upsilon_E$ dominates the instability growth-rate and the variations in the density-gradient scale length plays a negligible role in evolving the unstable modes especially in the range above 1 km. Moreover, for large values of electrons $\bm{{E}\times{B}}$ drifts only the variation in the density-gradient scale-length below 1 km has an effect on the unified instability growth-rate. Therefore, we expect to see no difference for all practical values of the background values of $L_n$ in the ionosphere during the daytime and nighttime.

Although the linear results show the roles that the density gradient scale-length and cross-field drift play in driving the dominant instabilities in the equatorial electrojet, these results do not show the saturation level of the electron density and the generation of effective Type-I 3-meter structures that  cannot be generated linearly. Therefore, nonlinear simulations are necessary to determine these characteristics of the equatorial electrojet instabilities.

\subsection{Nonlinear Results}\label{subsection:Nonlinear}
\begin{figure}
 \centering
 \noindent\includegraphics[width=1.0\textwidth]{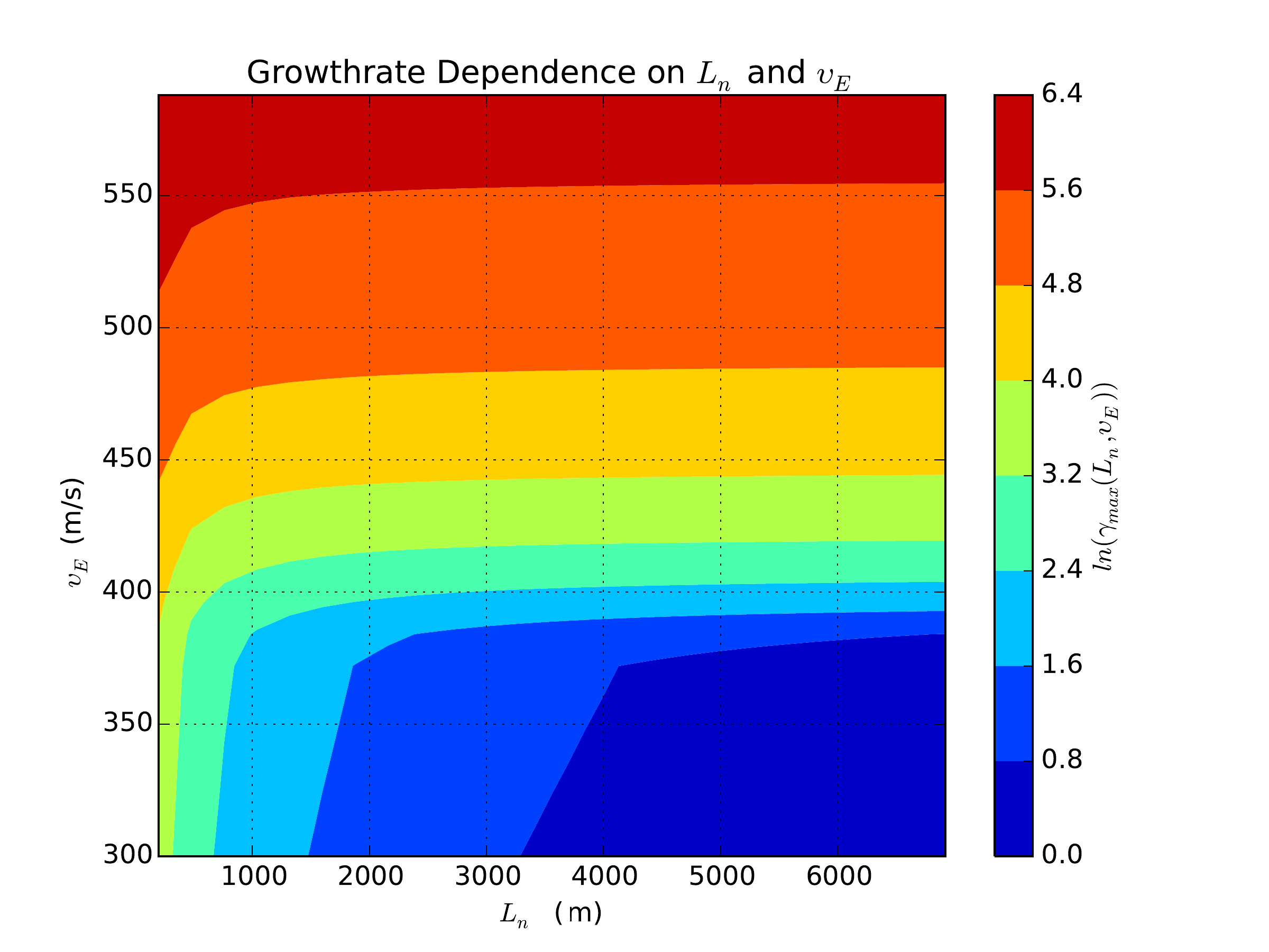}
 \noindent\caption{\small The dependence of the growth-rates at 105 km altitude on the density-gradient scale-length $L_n$ and the cross-field drift $\upsilon_E$. The large scale length does not have much effect on the growth-rate; for example,  above the ion-acoustic speed scale-lengths greater than 2 km have no effect on the growth-rate.}
 \label{fig:Growth2D}
\end{figure}

\begin{figure}
 \centering
 \begin{tabular}{c}
 \includegraphics[width=1.0\textwidth,height=0.7\textwidth]{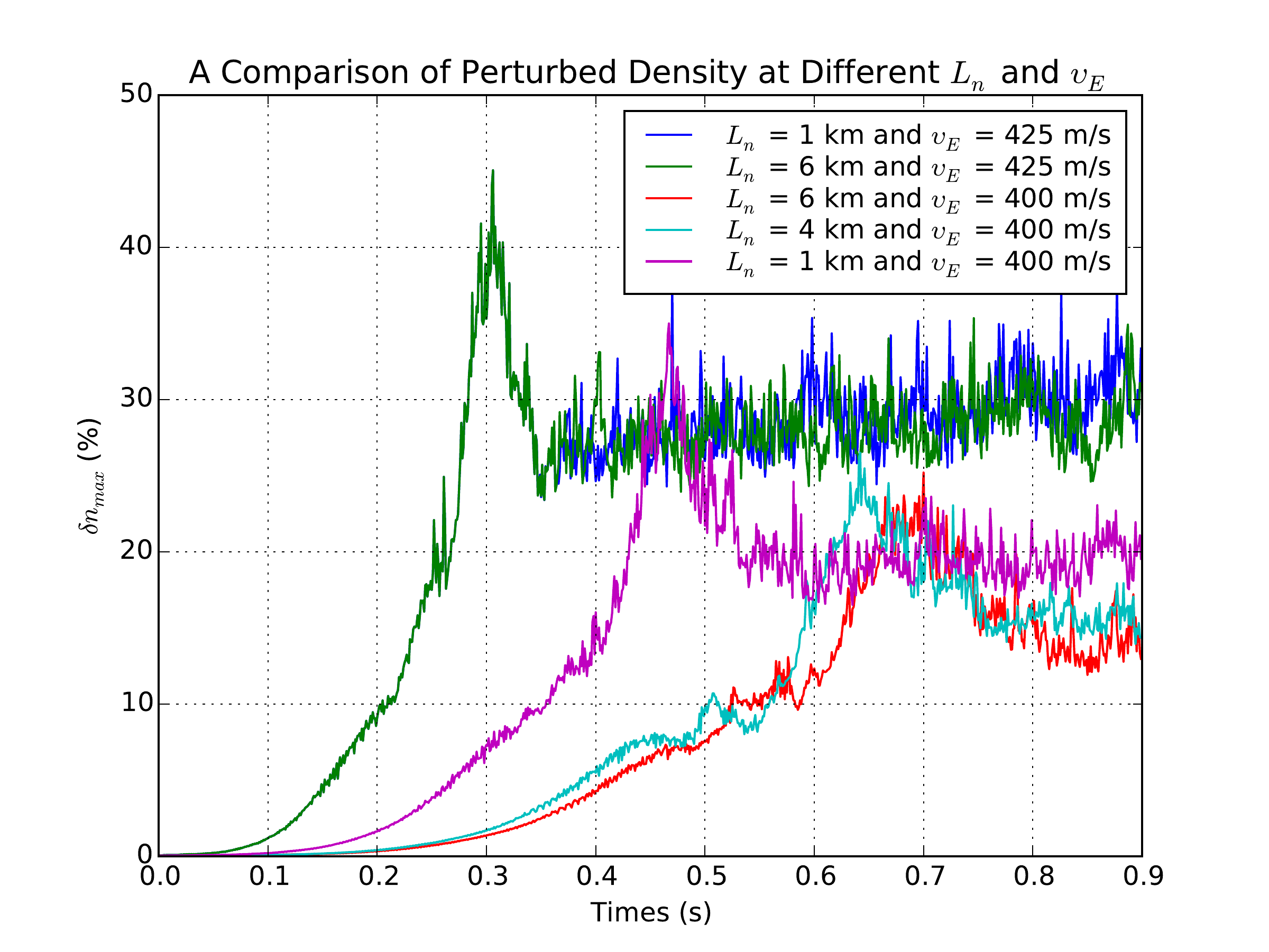}
 \end{tabular}
 \caption{\small A comparison between perturbed density maxima for different density-gradient scale-lengths, $L_n$ = 1, 4, and 6 (km) and cross-field ($\bm{{E}\times{B}}$) drifts, $\upsilon_E$ = 400 and 425 (m/s).} 
 \label{fig:NeMaxComp}
\end{figure}

Several nonlinear simulation results of the turbulence in the plasma density, perturbed electric field, and flux asymmetries in the unified model were presented by Hassan \textit{et al.}~[2015],  along with a comparison between the plasma turbulence during the solar maximum and solar minimum conditions. The simulation results showed three phases for the plasma turbulence, the growing phase where the linear terms dominate the plasma dynamics, the transition phase where the nonlinear terms have appreciable amplitude compared to the linear terms, and the saturation phase where the dynamics have reached a statistically steady state balance. These three phases in the simulation can be seen in the different study cases in Figure \ref{fig:NeMaxComp}.\\
\hfill\\
A comparison between the maximum perturbed electron density for different density-gradient scale-lengths is shown in Figure  \ref{fig:NeMaxComp},   where we can see that the time of the plasma density evolution in the linearly growing domain of the simulation is faster with shorter density scale-lengths. Although they are somewhat  different in the linear domain, the difference between the plasma density evolution for  the cases of $L_n$ = 4 km and 6 km is not  big  in the saturated state of the simulation.  However, a short scale-length of 1 km makes a big difference,  both in the linear growing mode of the perturbed density and in  the saturated state,  which shows an increase in the maximum fluctuation of the plasma density of more than 5\% compared to the background. The small difference between the $L_n$ = 4 and 6 km cases supports  the linear results of Figure \ref{fig:Growth2D},   where the variation in the gradient-density scale-length above 2 km has a  very small effect on the growth-rate for $\upsilon_E$ = 400 m/s.\\
\hfill\\
For the case of strong electron $\bm{{E}\times{B}}$ drift when $\upsilon_E$ = 425 m/s, different magnitudes of density-gradient scale-length ($L_n$ = 1 and 6 km) do not translate into differences in the growing mode of the simulation; however,  they match perfectly as   expected from Figure \ref{fig:Growth2D},  where the variation in $L_n$ from (approximately) 1 km gives rise to a negligible variation in the linear growth-rate under strong driving conditions of the Farley-Buneman instability. At saturation in the simulation, the difference is very small. On the other hand, an increase of the $\bm{{E}\times{B}}$ drift of 25 m/s with the same density-gradient scale-length causes a significant change in the saturation level of the perturbed electron density, it falling  in the range between 10 - 15 \%.\\
\hfill\\
Therefore, these simulation results show the importance of the density-gradient scale-length in driving the plasma gradient-drift instability in the absence of Farley-Buneman instability at small magnitudes of the electrons drift speed (below the ion-acoustic speed). However, in the presence of Type-I instability the role of the density gradient is negligible in driving the instability for scale-lengths larger than 2 km. Moreover, in the case of strong driving conditions with large cross-field drift,  the density gradient of any practical scale from 1 km does not play any role in driving the Farley-Buneman instability. This explains the disappear of the gradient-drift instability from the radar echoes in presence of the small structures formed in the presence of Farley-Buneman instability and energy cascading mechanism [Farley 2009].\\
\hfill\\
To conclude, the density-gradient and cross-field drift are co-players in driving Type-II instability in the absence of Type-I instability, but they dominate interchangeably at different ranges of values of $L_n$ and $\upsilon_E$. Above the ion-acoustic speed there is a transition region where both the density-gradient and cross-field drift play important roles in driving both types of instabilities, but at large values of cross-field drifts ($\upsilon_E\gg$ 400 m/s) the role of density-gradient in driving the instabilities is too small to be considered and the Type-I instability dominates and its dynamics are controlled by the electron $\bm{{E}\times{B}}$ drift. The study, described below, of the energy distribution of plasma fluctuations of different scales and the forward and reverse cascade of energy among these structures will shed some light on the generation of short and long wavelength irregularities in the equatorial electrojet.

\section{Energy Components and Transfer in the Unified Fluid Model}\label{section:EnergyConservation}
\begin{figure}
 \centering
 \begin{tabular}{c}
 \includegraphics[width=0.6\textwidth,height=0.7\textwidth]{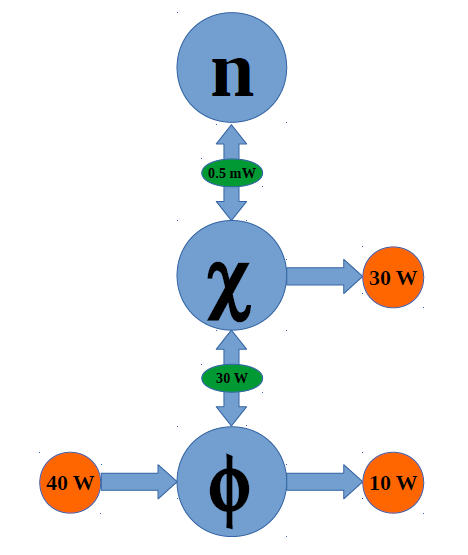}
 \end{tabular}
 \caption{\small The input source of energy from the boundaries in the electric potential dynamical field and the dissipation of energy in the electrons and ions collisions with the background. Also, the energy coupling between the evolving fields is identified.} 
 \label{fig:EnergyDisChart}
\end{figure}

In the linear and nonlinear results of  Figures \ref{fig:Growthrates},  \ref{fig:Growth2D},  and \ref{fig:NeMaxComp},   we showed the effect of different scale-lengths of density-gradient and electric-potential-gradient on the linear growth-rate of the unstable modes and the evolution of the perturbed quantities over the growing, transition, and saturation regions of the nonlinear simulation. This demonstrates that  available free energy is injected into the system from the vertical boundaries through  constant gradients of the plasma density $\partial_zn_o$ and electric potential $\partial_z\phi_o$,  with their importance in driving the instabilities varying depending on their local values in the ionosphere. In contrast, the energy is dissipated  out of the system due to the electron and ion viscosities,  where their collision frequencies with the background neutral particles acting as energy sinks.\\
\hfill\\
Therefore, by identifying and separating out the source, dissipation, and dynamical  components of the governing  unified fluid model of equations  (\ref{eq:Uni3fIonContinuity}) -- (\ref{eq:Uni3fQuasineutrality}), we are poised to trace energy transfer.  To this end we  rewrite the unified fluid model  in the following fully nonlinear form:
\begin{eqnarray}\label{eq:IonContinuitySeparate}
\hspace{1.5 cm}
\partial_t{\delta{n}} &=& \bm\nabla\cdot(\delta{n}\bm\nabla{\delta\chi}) + \bm\nabla\cdot(n_o\bm\nabla{\delta\chi})
\\
\label{eq:IonMomentumSeparate} 
\partial_t{\nabla^2\delta\chi} &=& \frac{\Omega_{ci}}{B}\nabla^2\delta\phi + \upsilon^2_{t_i}\nabla^2\delta{n}+\frac{1}{2}\nabla^2|\bm\nabla\delta\chi|^2
\\
&& \qquad 
- \nu_{in}\nabla^2\delta\chi+\frac{4}{3}\frac{\upsilon^2_{t_i}}{\nu_{in}}\nabla^4\delta\chi
\nonumber\\
 \label{eq:QuasiNeutralitySeparate}
\partial_t\nabla^2\delta\phi &=& -B\Omega_{ce}n_o^{-1}\bm{\nabla}\cdot(\delta{n}\bm{\nabla}\delta\chi)-B\Omega_{ce}n_o^{-1}\bm{\nabla}\cdot(n_o\bm{\nabla}\delta\chi)
\\
&& \qquad 
-\ \Omega_{ce}[\phi_o,\ln(n_o)]-\Omega_{ce}[\delta\phi,\ln(n_o)]-\Omega_{ce}[\phi_o,\delta{n}]-\Omega_{ce}[\delta\phi,\delta{n}]
\nonumber\\
&& \qquad 
- \frac{1}{B}[\phi_o,\nabla^2\delta\phi]-\frac{1}{B}[\delta\phi,\nabla^2\delta\phi]
\nonumber\\
&& \qquad 
+ \ \frac{T_e\nu_{en}}{e}\nabla^2\ln(n)+\frac{T_e\nu_{en}}{e}\bm{\nabla}\ln(n)\cdot\bm{\nabla}\ln(n)
\nonumber\\
&& \qquad 
-\ \nu_{en}\nabla^2\phi-\nu_{en}\bm{\nabla}\ln(n)\cdot\bm{\nabla}\phi.
\nonumber
\end{eqnarray}
The physics of the ionosphere plasma in the E-region is such that the energy in the  system is contained  in the kinetic energy of the magnetized electrons due to the $\bm{\delta{E}\times{B}}$ drifts, the kinetic energy of the unmagnetized ions, and the internal thermal energy in the plasma species. Therefore, the total energy of these three components, respectively, can be written in the following form:
\begin{align}\label{eq:SysEnergy}
E_T = \int d^2x' \left(
                  \frac{n_om_e}{2B^2}|\bm{\nabla}\delta\phi|^2+\frac{m_i}{2}\delta{n}|\bm{\nabla}\delta\chi|^2+\frac{1}{2}m_i\upsilon^2_{t_i}\delta{n}^2
            \right).
\end{align}
The total energy in equation (\ref{eq:SysEnergy}) is shown  in Hassan \textit{et al.}~[2016] to serve as the Hamiltonian of a noncanonical Hamiltonian system [Morrison~1998].\\
\hfill\\
This total energy reaches a saturated state, i.e.,  when there is zero time-averaged rate of change because of  a balance between the energy sources at the system boundaries and the energy sink represented by the electron and ion  viscosities, which as noted  model their  corresponding collisions with the background neutrals. The rate of change of  the total energy can be written in terms of the energy source (S), the energy dissipation (D), the nonlinear energy transfer between different scales of plasma structures (N), and the energy coupling between the evolving fields in the system (C) as  follows:
\begin{eqnarray}\label{eq:EnergyDotItems}
\dot{E}_T &=& S + D + N + C
\nonumber\\
  &=& \{S + D + N + C\}_{n} + \{S + D + N + C\}_{\chi} + \{S + D + N + C\}_{\phi},
  \end{eqnarray}
where in the second equality the subscripts represent the contributions from each of equations (\ref{eq:IonContinuitySeparate})--(\ref{eq:QuasiNeutralitySeparate}), respectively.  To study the physics in the source, sink, nonlinear, and coupling terms of the three evolving fields $(\delta{n}, \delta\phi, \delta\chi)$ in equation (\ref{eq:EnergyDotItems}), we need to find the rate of change of the total energy   (\ref{eq:SysEnergy}),  which can be found from  the following:
\begin{eqnarray}
\dot{E} &=& \int d^2x' \Big(
-\frac{m_en_o}{B^2}\delta\phi\partial_t\nabla^2\phi
   +\frac{m_i}{2}|\bm{\nabla}\chi|^2\partial_t{\delta{n}}
   \nonumber\\ 
 &&
 \hspace{4cm} -\ m_i\partial_t\delta\chi\bm{\nabla}\cdot(n\bm{\nabla}\delta\chi)  
 + m_i\upsilon^2_{t_i}\delta{n}\partial_t{\delta{n}}
       \Big), 
 \label{eq:EDot}
\end{eqnarray}
where we used integration by parts, 
\[
\int d^2x' \bm{\nabla} {f}\cdot \partial_t\bm{\nabla} {g} = \int d^2x' \bm{\nabla}\cdot\left({f}\partial_t\bm{\nabla}{g}\right) - \int d^2x' {f}\partial_t\nabla^2 {g} = - \int d^2x' {f}\partial_t\nabla^2 {g},
\]
in the first and second terms of  the integral, which requires either Dirichlet or periodic boundary conditions for all integrated evolving fields represented by $f$ and $g$, and the periodic boundary condition is employed in the simulation of the dynamical equations in the unified fluid model.\\
\hfill\\
Then, from the set of partial differential equations (\ref{eq:IonContinuitySeparate}), (\ref{eq:IonMomentumSeparate}), and (\ref{eq:QuasiNeutralitySeparate}) that governs the plasma dynamics in the equatorial electrojet instabilities with the energy balance of  equation (\ref{eq:EDot}), we can rewrite the rate of energy change of the system in the following form:\\
\begin{align}\label{eq:EnergyDot}
\begin{split}
\dot{E}_T &= \int d^2x'\Big( e\delta\phi\left(\bm{\nabla}\cdot(n_o\bm{\nabla}\delta\chi)+\bm{\nabla}\cdot(\delta{n}\bm{\nabla}\delta\chi)\right)\\ 
        &\qquad +   \frac{en_o}{B}\delta\phi[\delta\phi,\ln(n_o)]+[\phi_o,\delta\tilde{n}]+[\delta\phi,\delta\tilde{n}]\\
        &\qquad  +  \frac{en_o}{B^2\Omega_{ce}}\delta\phi\left([\phi_o,\nabla^2\delta\phi]+[\delta\phi,\nabla^2\delta\phi]\right)\\
        & \qquad -   en_o\rho_e^2\nu_{en}\delta\phi
                      \left(    
                             \nabla^2\delta\tilde{n} + 2\bm{\nabla}\ln(n_o)\cdot\bm{\nabla}\delta\tilde{n} +\bm{\nabla}\delta\tilde{n}\cdot\bm{\nabla}\delta\tilde{n}
                      \right)\\
        & \qquad  +   \frac{en_o\nu_{en}}{B\Omega_{ce}}\delta\phi
                      \left(    
  \nabla^2\delta\phi+\bm{\nabla}\ln(n_o)\cdot\bm{\nabla}\delta\phi + \bm{\nabla}\delta\tilde{n}\cdot\bm{\nabla}\phi_o+\bm{\nabla}\delta\tilde{n}\cdot\bm{\nabla}\delta\phi
                      \right)\\
        &\qquad +  m_i\nu_{in}\delta\chi\left(\bm{\nabla}\cdot(n_o\bm{\nabla}\delta\chi)+\bm{\nabla}\cdot(\delta{n}\bm{\nabla}\delta\chi)\right)\\
        &\qquad -   \frac{4}{3}\frac{m_i\upsilon_{t_i}^2}{\nu_{in}}\nabla^2\delta\chi\left(\bm{\nabla}\cdot(n_o\bm{\nabla}\delta\chi)+\bm{\nabla}\cdot(\delta{n}\bm{\nabla}\delta\chi)\right)\\
        &\qquad -  \left(                     e\delta\phi+m_i\upsilon^2_{t_i}\delta{n}\right)\left(\bm{\nabla}\cdot(n_o\bm{\nabla}\delta\chi)+\bm{\nabla}\cdot(\delta{n}\bm{\nabla}\delta\chi)
                      \right)\\
        &\qquad +   m_i\upsilon^2_{t_i}\delta{n}\left(\bm{\nabla}\cdot(n_o\bm{\nabla}\delta\chi)+\bm{\nabla}\cdot(\delta{n}\bm{\nabla}\delta\chi)\right)\Big). 
\end{split}					
\end{align}
A close look at the physics in equation (\ref{eq:EnergyDot}) leads to  the understanding of the contribution of each term to the rate of change of the total energy of  (\ref{eq:EnergyDotItems}), thereby  allowing  us to identify  the special roles  of each  in the system dynamics.  In this way we identify terms as  source (S), dissipation (D),  energy transfer (N) of each evolving field,  and  energy coupling (C) between the evolving fields. Therefore, we can write their contributions explicitly as follows:\\
\begin{align}\label{eq:EnergyContributions}
\begin{split}
S_{\phi} &= \int d^2x'\, \frac{en_o}{B}\delta\phi\left([\delta\phi,\ln(n_o)]+[\phi_o,\delta{n}] + \frac{1}{B\Omega_{ce}}[\phi_o,\nabla^2\delta\phi]\right)\\
D_{\phi} &= \int d^2x'\, \Big(\frac{en_o\nu_{en}}{B\Omega_{ce}}\delta\phi
                            \left(\nabla^2\delta\phi +\bm{\nabla}\ln(n_o)\cdot\bm{\nabla}\delta\phi +\bm{\nabla}\delta{n}\cdot\bm{\nabla}\phi_o +\bm{\nabla}\delta{n}\cdot\bm{\nabla}\delta\phi\right)\\
               & \qquad -\  en_o\rho_e^2\nu_{en}\delta\phi
                            \left(\nabla^2\delta{n} + 2\bm{\nabla}\ln(n_o)\cdot\bm{\nabla}\delta{n}+\bm{\nabla}\delta{n}\cdot\bm{\nabla}\delta{n}\right)\Big)
 \\
N_{\phi} &= \int d^2x'\,  \frac{en_o}{B}\delta\phi\left([\delta\phi,\delta{n}] +\frac{1}{B\Omega_{ce}}[\delta\phi,\nabla^2\delta\phi]\right)\\
C_{\phi\chi} &= \int d^2x'\,  e\delta\phi\left(\bm{\nabla}\cdot(n_o\bm{\nabla}\delta\chi) +\bm{\nabla}\cdot(\delta{n}\bm{\nabla}\delta\chi)\right)\\
D_{\chi} &= \int d^2x'\Big( m_i\nu_{in}\delta\chi\left(\bm{\nabla}\cdot(n_o\bm{\nabla}\delta\chi)+\bm{\nabla}\cdot(\delta{n}\bm{\nabla}\delta\chi)\right)\\
         &\qquad - \  \frac{4}{3}\frac{m_i\upsilon_{t_i}^2}{\nu_{in}}\nabla^2\delta\chi\left(\bm{\nabla}\cdot(n_o\bm{\nabla}\delta\chi)+\bm{\nabla}\cdot(\delta{n}\bm{\nabla}\delta\chi)\right)\Big)
         \\
C_{\chi\phi} &= -\int d^2x'\,  e\delta\phi\left(\bm{\nabla}\cdot(n_o\bm{\nabla}\delta\chi) +\bm{\nabla}\cdot(\delta{n}\bm{\nabla}\delta\chi)\right)\\
C_{\chi{n}} &= -\int d^2x'\,  m_i\upsilon^2_{t_i}\delta{n}\left(\bm{\nabla}\cdot(n_o\bm{\nabla}\delta\chi) +\bm{\nabla}\cdot(\delta{n}\bm{\nabla}\delta\chi)\right)\\
C_{n\chi} &= \int d^2x'\,  m_i\upsilon^2_{t_i}\delta{n}\left(\bm{\nabla}\cdot(n_o\bm{\nabla}\delta\chi) +\bm{\nabla}\cdot(\delta{n}\bm{\nabla}\delta\chi)\right). 
\end{split}					
\end{align}
Thus,  energy is always injected into the system from the dynamics of the electric potential, $S_{\phi}$, and there is no other energy source can be recognized in the dynamics of the ion  velocity potential or plasma density evolving fields. Part of the energy injected into the system is dissipated internally via  the electron viscosity, $D_{\phi}$; however,  the other part is transferred to the ions velocity potential dynamical equation through the coupling term, $C_{\phi\chi}$, between the fluctuating electric potential and velocity potential fields  that will  be dissipated in the collision process of the ions with the background neutral particles, $D_{\chi}$. A chart of the amount of energy injected/dissipated/coupled in/from/between the fields is shown in Figure \ref{fig:EnergyDisChart} --  more discussion will be  provided in section (\ref{subsection:EnergySourcesSinks}) where  the physical mechanisms of injecting and dissipating energy for  the system are discussed.\\
\hfill\\
From the coupling terms in equation (\ref{eq:EnergyContributions}), we can see that the third term in equation (\ref{eq:EDot}) plays the coupling role between the ion velocity potential and the other evolving fields (it also couples with itself). This happens because this term represents conservation in the plasma density in addition to the quasineutrality assumption that is employed in the system. Surprisingly, the dynamics of the ion  velocity potential and density are coupled together through a coupling term, $C_{n\chi}$, which is the only term in the continuity equation of the plasma density. This explains the tying of the total energy in the ion velocity potential with the  plasma density that is found in Figure \ref{fig:Etotal_T_Ln6000Vexb400}  and bears  on the growing and transitional  regimes of the simulations. It is also noticed that there is no direct coupling between the perturbed plasma density $\delta{n}$ and the perturbed electric potential $\delta\phi$, and all the coupling dynamics between these two fields has to happen through the cross-coupling of these evolving fields and the perturbed ion velocity potential $\delta\chi$.\\

\subsection{Tracking the Total Energy in Simulations}\label{subsection:TotalEnergy}
\begin{figure}
 \centering
 \begin{tabular}{c}
 \includegraphics[width=1.0\textwidth,height=0.7\textwidth]{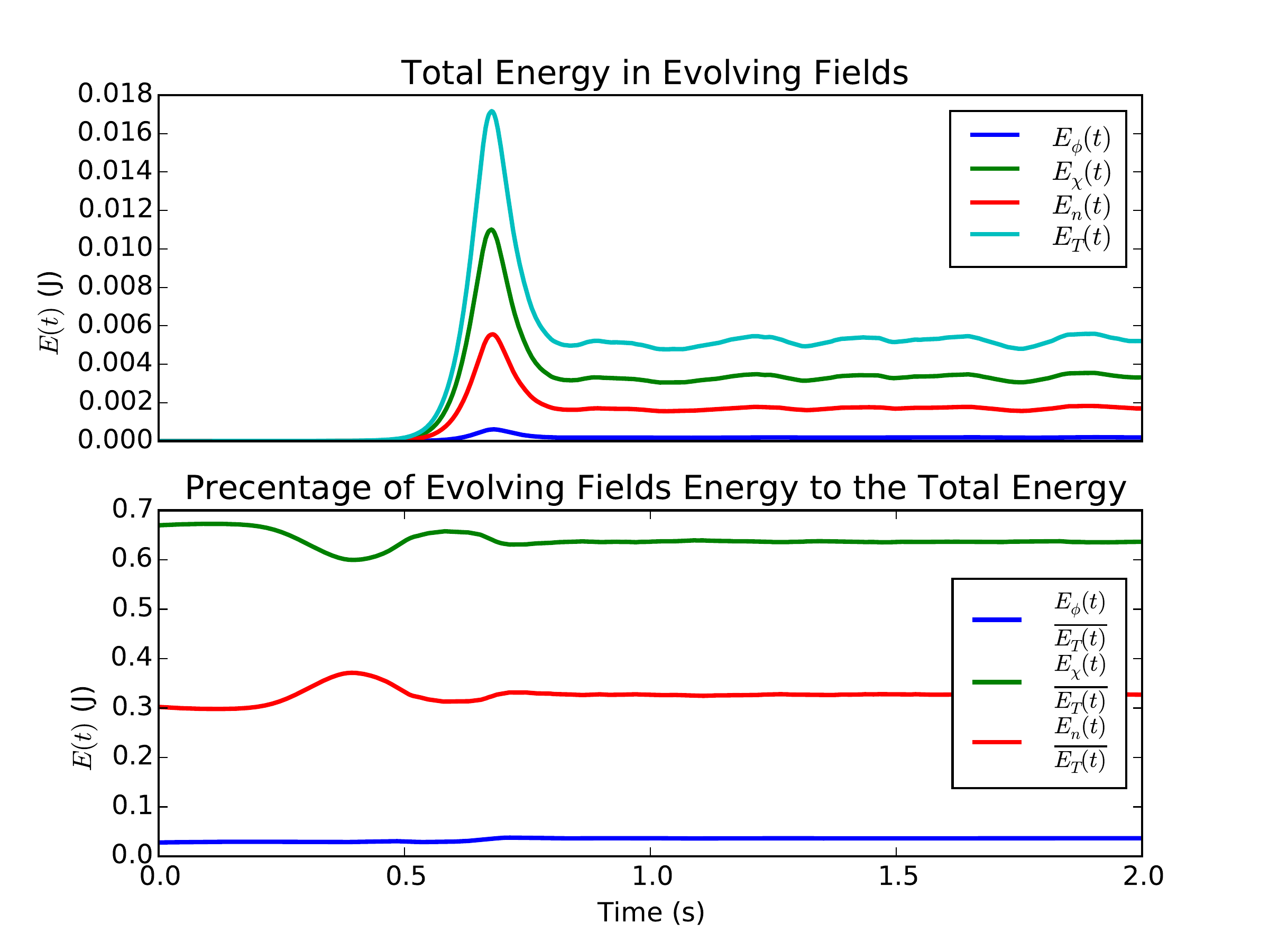}
 \end{tabular}
 \caption{\small The total energy in the evolving fields  $(\delta{n},\delta\phi,\delta\chi)$ (top panel) and the ratio of the energy in each evolving field to the total energy (bottom panel) for $L_n$ = 6 km and $\upsilon_E$ = 400 m/s. This shows the role that the ions playing in evolving the system as the field that contains the largest portion of energy in dynamical system.} 
 \label{fig:Etotal_T_Ln6000Vexb400}
\end{figure}

\begin{figure}
 \centering
 \begin{tabular}{c}
 \includegraphics[width=1.0\textwidth,height=0.7\textwidth]{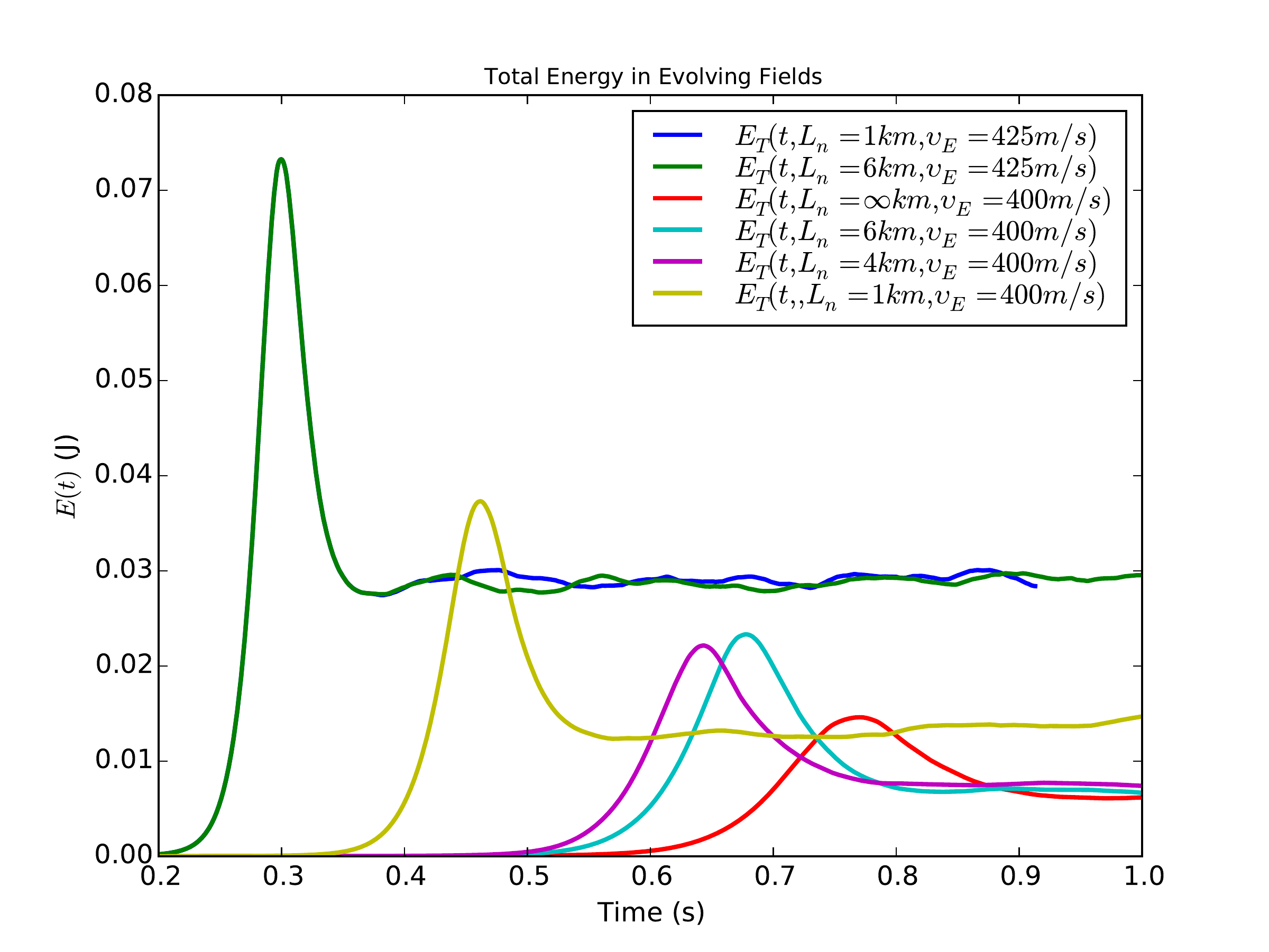}
 \end{tabular}
 \caption{\small A comparison between the total energy (the sum of the ion and electron kinetic energy and the plasma internal energy) for different magnitudes of the density scale-length $L_n$ = 1, 4, 6, and $\infty$ (km) and  drifts $\upsilon_E$ = 400 and 425 (m/s).} 
 \label{fig:EtotalCompare}
\end{figure}

\begin{figure}
 \centering
 \begin{tabular}{c}
 \includegraphics[width=1.0\textwidth,height=0.6\textwidth]{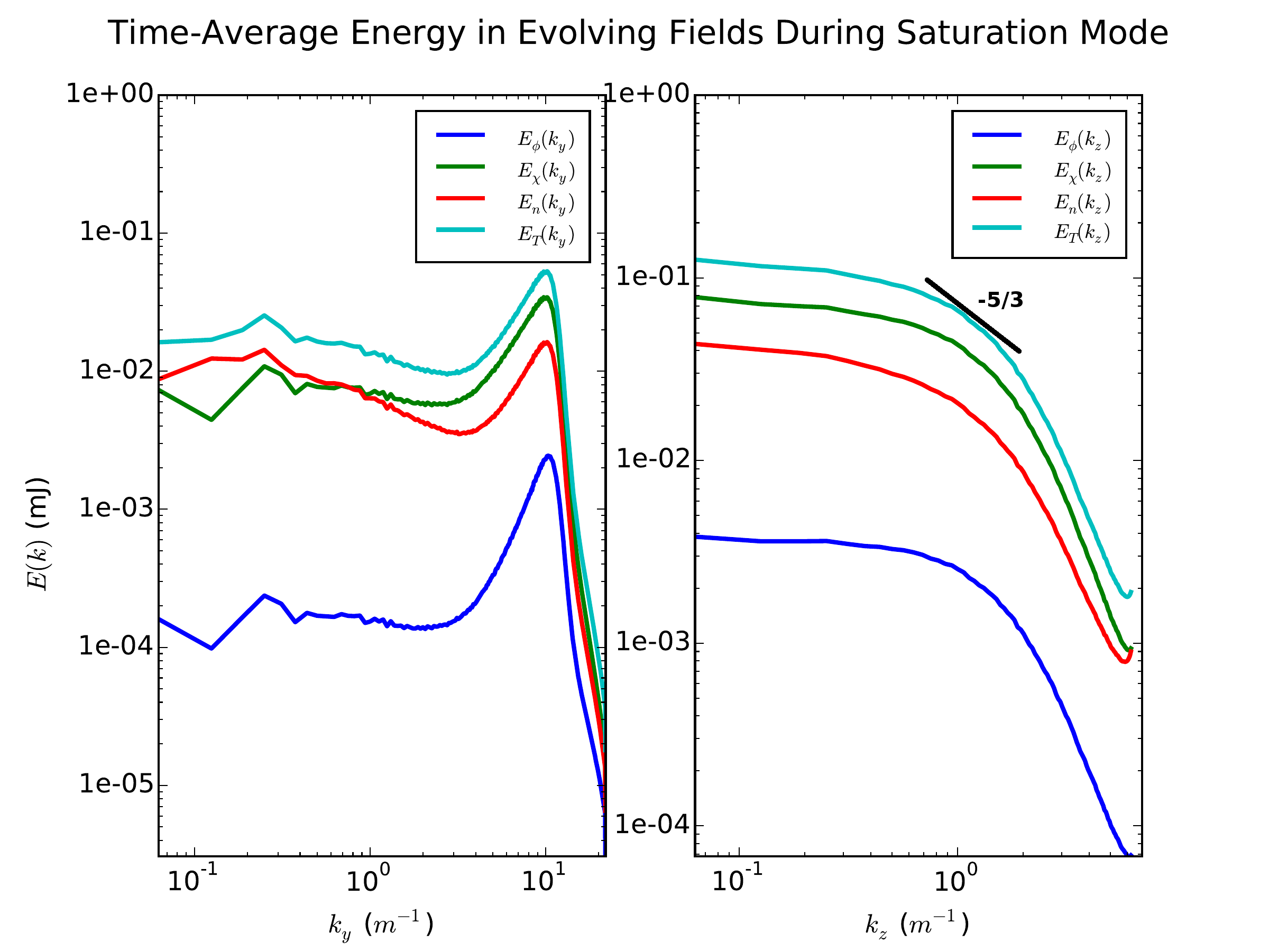}
 \end{tabular}
 \caption{\small The time-average of the energy in the evolving fields  $(\delta{n},\delta\phi,\delta\chi)$ and their corresponding total energy during  the saturation state of the simulation for $L_n$ = 6 km and $\upsilon_E$ = 400 m/s as a function of the horizontal ($k_y$) (left panel) and vertical ($k_z$)  (right panel) wavenumbers.} 
 \label{fig:Etotal_Ky_Ln6000Vexb400}
\end{figure}

As noted before,  the total energy of  (\ref{eq:SysEnergy}) is composed of  three types, the kinetic energy of ions and electrons and the internal thermal energy of the plasma species. The total energy $E_{T}$ and the ratio between  each of the three energy components and the total energy are shown in Figure~\ref{fig:Etotal_T_Ln6000Vexb400} for the case where  $L_n$ = 6 km and $\upsilon_E$ = 400 m/s. Although it strongly controls the system dynamics through the temporal variations of the electric potential, the electron  kinetic energy $E_{\phi}$ due to the $\bm{\delta{E}\times{B}}$ drifts makes a very small contribution to the total energy because it dependends  on the small-amplitude fluctuation of the perturbed electric potential and the small inertia of the electrons.  On the other hand, the kinetic energy content of the heavy ions $E_{\chi}$, depending on both the density and ion velocity potential,  has almost double the internal thermal energy $E_n$ in the system. The sum of $E_{\chi}$ and $E_{n}$ amounts to   approximately 98\% of  the total energy in the dynamical  system, as shown in the top-panel of Figure~\ref{fig:Etotal_T_Ln6000Vexb400}.\\
\hfill\\
The ratio between the energy in the evolving fields and the total energy in the system is shown in the bottom-panel of Figure~\ref{fig:Etotal_T_Ln6000Vexb400}. This ratio shows that the plasma internal thermal energy and ion  kinetic energy are partially exchanging   energy  over the growing and transition phases of the simulation.  Then,  their energy content becomes and remains almost flat  during the saturation state. The coupling between the energy components in the ions shown in equation (\ref{eq:EnergyContributions}) explains the strong tying of the dynamics in the plasma density and ions velocity potential. The interrelation between the energy content in the plasma density, ion velocity potential, and electron potential has been calculated in other cases with  different density-gradient scale-lengths $L_n$ and cross-field drifts $\upsilon_E$: the energy content of the ion velocity potential is always larger than that of the plasma density, and the perturbed electric potential has the smallest energy content in the system.\\
\hfill\\
A comparison between the total energy profiles for different magnitudes of the plasma density scale-length ($L_n$ = 1, 4, 6, and $\infty$ km) and the cross-field drifts ($\upsilon_E$ = 400 and 425 m/s) is shown in 
Figure~\ref{fig:EtotalCompare}. For an electron drift of magnitude $\upsilon_E$ = 400 m/s, the increase in the density scale-length gives rise to an increase of the rate of evolution of the energy in the growing and transition states of the simulation and a different level of the total energy at the saturation region. However, for $\upsilon_E$ = 425 m/s, the effect of the density scale-length is negligible over all phases of energy evolution in the system. The saturation energies in the cases of $L_n$ = 4, 6, and $\infty$ km are almost the same,  but it takes more time to reach almost the same level of saturation as the density-gradient scale-length increases. The energy at the saturated  state of the simulation almost doubles from $\sim$7 mJ when $L_n$ = 4, 6, and $\infty$ km to $\sim$14 mJ when $L_n$ = 1 km. Similarly, the energy level almost doubles when the $\bm{{E}\times{B}}$ drift increases from $\upsilon_E$ = 400 m/s to $\upsilon_E$ = 425 m/s.\\
\hfill\\
These results underscore the fact that the density scale-length $L_n$ and the cross-fields drift $\upsilon_E$ are the energy sources in our dynamical system, and that they play important roles in driving the system in its linear growing mode phase and ultimate termination  at different saturation levels.
While a small increase in the cross-field drift (from $\upsilon_E$ = 400 m/s to $\upsilon_E$ = 425 m/s) causes a dramatic increase in the growth rate of the unstable modes and  results in a larger amplitude of the fluctuating density,   changes in the density-gradient scale-length do not show the same effect,  especially for the cases of $L_n$ = 4, 6, $\infty$ km. 
However, when the scale-length of the density-gradient approaches the effective size of  the simulation box ($L_n$ = 1 km), we noticed a large change in the growth rate of the unstable modes and the maximum perturbed density, but one that  is still not comparable  to the change induced by the cross-field drifts $\upsilon_E$. Therefore, these results again show the effect of different combinations of the density-gradient scale-length $L_n$ and cross-field drift speed $\upsilon_E$ on  driving the instabilities in the system are similar to the linear and nonlinear results described in subsections \ref{subsection:Linear} and \ref{subsection:Nonlinear}.\\
\hfill\\
The time-average of the total energy and its components as a function of the horizontal wavenumber $k_y$ and vertical wavenumber $k_z$  during the saturated state of the simulation is shown in Figure~\ref{fig:Etotal_Ky_Ln6000Vexb400},  for $L_n$ = 6 km and $\upsilon_E$ = 400 m/s. The two peaks at small and large wavenumbers that were  seen in the linear growth-rate versus  $k_y$  are still found during the nonlinear saturated state as shown in the left-panel of Figure~\ref{fig:Etotal_Ky_Ln6000Vexb400}. The peak at the small-wavenumber is still around 0.1 $m^{-1}$,  similar to the corresponding peak in the linear growth-rate.  However, the peak at the large-wavenumber has shifted from $\sim 7-8 m^{-1}$ to $\sim 10-11 m^{-1}$.  This suggests that the nonlinear dynamics of the system transfers  energy to structures of smaller wavelength, which would  explain  the strong backscattering of radar echoes that come from the short structures  compared to long ones and the disappearance of the Type-II instability in the presence of Type-I instability, and agrees with the large growth-rate of small-scale structures compared to the growth-rate of the unstable waves of long-wavelength in the linear regime. It is also found that the size of the small plasma structures that have the largest energy content are  strongly dependent on the magnitudes of the density-gradient scale-length $L_n$ and cross-field drifts ($\upsilon_E$).\\
\hfill\\
The time-average of the energy in the vertical direction of  $k$-space during the saturation mode is shown in the right-panel of Figure~\ref{fig:Etotal_Ky_Ln6000Vexb400}. The amount of energy content in the small structures is one order of magnitude larger than the amount of energy content in the corresponding structures in the horizontal direction. According to this scale-dependent energy content, one would  expect to observe strong echoes from the vertical structures of 1 - 5 meters scale-size and not from the horizontal ones of the same scale, which is consistent with the recorded radar observations [Kudeki \textit{et al.}~1987] and sounding rocket measurements [Pffaf \textit{et al.}~1987a,b] during the CONDOR campaign. Moreover, while no explicit power-law can be found in the horizontal and vertical k-space, these results also suggest the presence of a forward energy cascading mechanism to transfer the energy from large plasma structures into  small ones, which will be discussed further in section  \ref{section:EnergyCascades}.

\subsection{Energy Sources and Dissipations}\label{subsection:EnergySourcesSinks}
\begin{figure}
 \centering
 \begin{tabular}{c}
 \includegraphics[width=1.0\textwidth,height=1.0\textwidth]{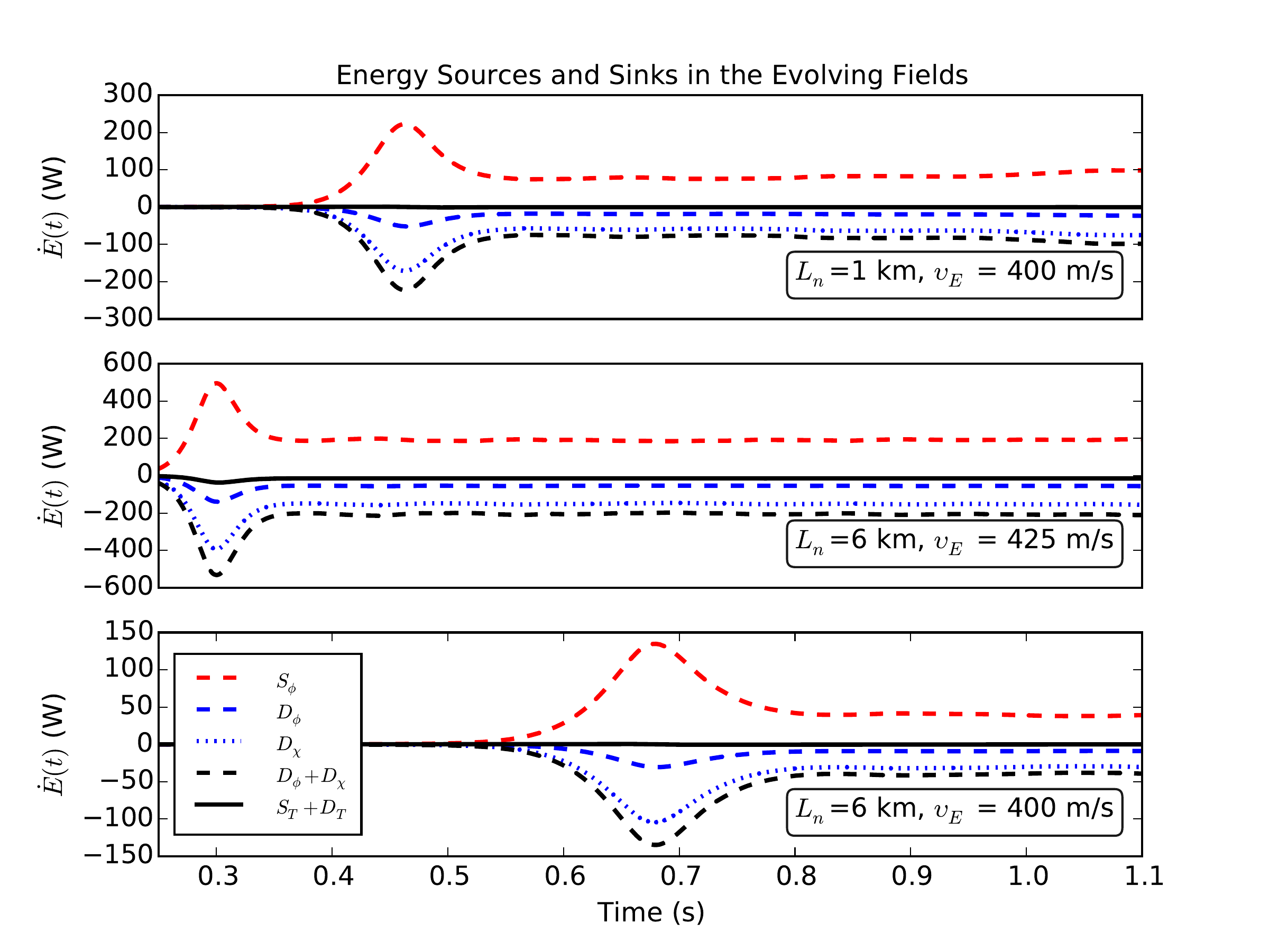}
 \end{tabular}
 \caption{\small The total rate of change of energy   sources and sinks for the evolving fields over the transition and saturation phases of the simulation for the  case $L_n$ = 1 and 6 km and $\upsilon_E$ = 400 and 425 m/s. This shows the dependence of the beginning of the transition and saturation regions of the simulation on the magnitude of the energy sources ($L_n$ and $\upsilon_E$).} 
 \label{fig:SDTComp}
\end{figure}

\begin{figure}
 \centering
 \begin{tabular}{c}
 \includegraphics[width=1.0\textwidth,height=0.7\textwidth]{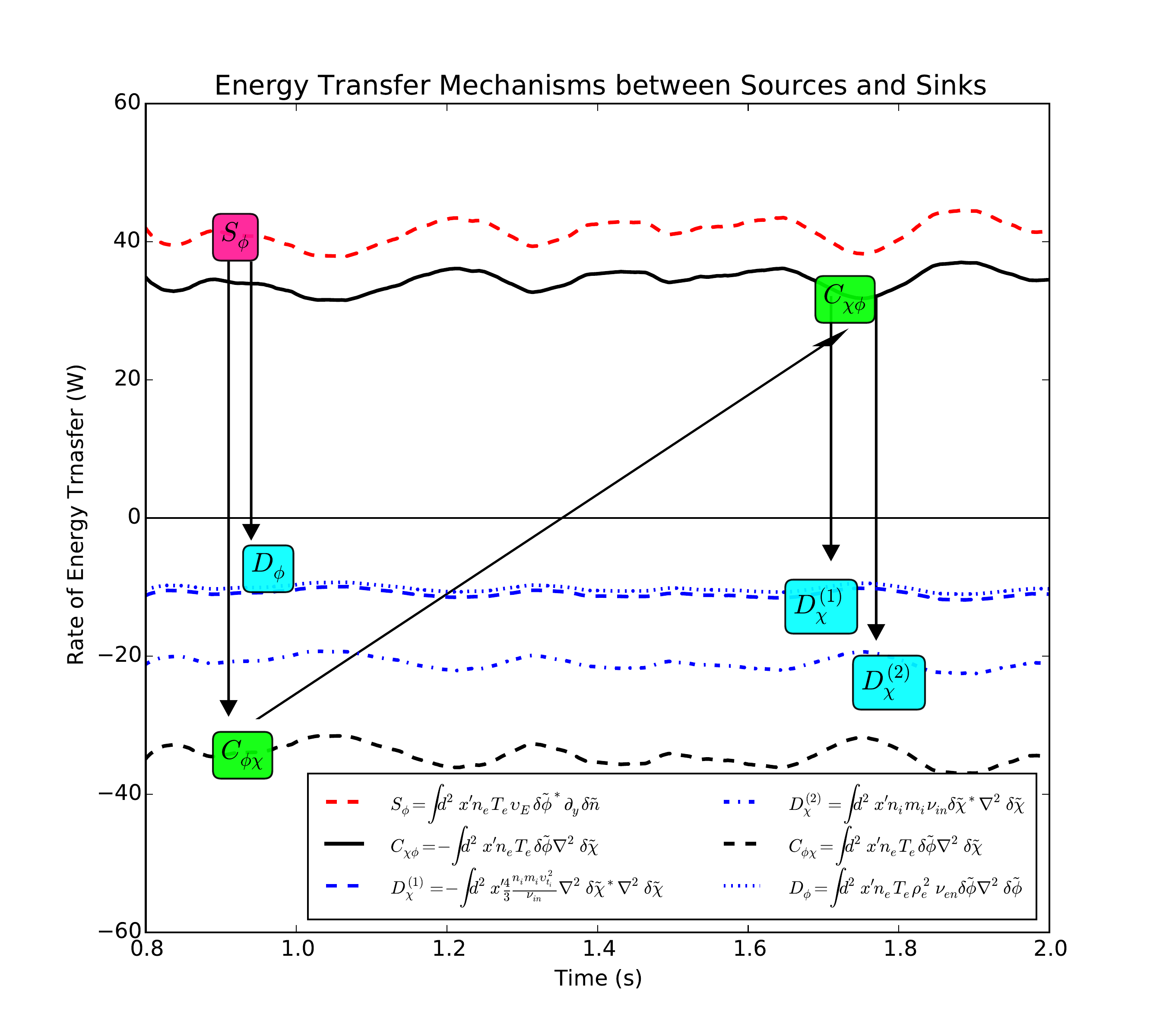}\\
 \end{tabular}
 \caption{\small The energy transfer mechanism and physics between the source and sink terms over the saturation state of the simulation for a case of $L_n$ = 6 km and $\upsilon_E$ = 400 m/s and the coupling between the fields. The thermal internal energy in the density field is not shown here because it is only coupled with the ions velocity potential and it has a very small magnitude.} 
 \label{fig:SrcDis_T_Ln6000Vexb400}
\end{figure}

Although studying the rate of energy transfer in each of the evolving fields $(\delta{n}, \delta\phi, \delta\chi)$ in equations (\ref{eq:EDot}) verifies total  energy conservation of the system, it does not show the role each field plays as source, sink, cross-field coupling, or inner-field coupling, and the various  redistribution mechanisms of components of the total energy.\\
\hfill\\
In Figure~\ref{fig:SDTComp} we show the various sources and dissipation of the energy in the ions and electrons for different magnitudes of the  density-gradient scale-length ($L_n$ = 1 and 6 km) and cross-field drift ($\upsilon_E$ = 400 and 425 m/s). In all cases of the simulation,  the major source of energy input for the unified fluid mode  is the electron kinetic energy, which also dissipates a small part of that energy via  collisions of the electrons with neutrals.  However, the heavy ions are responsible for most of the dissipation of energy in the  system because of their large collision frequency with the neutral particles in the atmosphere.  In our representation of energy sources and dissipation,  we do not include the coupling terms between the evolving fields because  they cancel each other,  as  can be shown from equation (\ref{eq:EnergyDot}).\\
\hfill\\
In all the  cases of Figure~\ref{fig:SDTComp},  we found that the total dissipation in the ions and electrons due to collisions is equal to the total energy coming into the electron dynamics from the sources, the gradients of the background density and electric potential in the vertical direction.  So,  we have determined  that the entire system satisfies energy conservation between the three evolving fields, as represented by the black solid line that passes through the $x$-axis at $\dot{E}_T$ = 0.\\
\hfill\\
Two more examples for the magnitudes of the energy sources and dissipation for different values of density scale-length ($L_n$ = 6 km) and cross-fields drift speed ($\upsilon_E$ = 425 and 400 m/s) are shown in the second and third panels of Figure~\ref{fig:SDTComp}, respectively. A comparison between the results in the three panels elucidates the significant effect of any small change in the $\bm{{E}\times{B}}$ drifts on the evolution of the equatorial electrojet instabilities.  This is  due to the availability of a large amount of free energy in the system. A similar but smaller effect takes  place with a short density-gradient scale-length ($L_n$ = 6 km vs $L_n$ = 1 km). This means that our system drives the Farley-Buneman (Type-I) instability stronger than the gradient-drift (Type-II) instability when the cross-field drift is much larger than the magnitude of the ion-acoustic speed in the local ionosphere.\\
\hfill\\
We further investigated the behavior of each term in the energy equation in conserving the energy in the system. We included only the terms that add, couple, or dissipate appreciable amounts of energy, dropping very small terms that are at least three orders of magnitude smaller than those kept.\\
\hfill\\
Figure~\ref{fig:SrcDis_T_Ln6000Vexb400} shows that the major source of energy comes from 
\[
S_{\phi} =  \int\! d^2x\,  n_e T_e \upsilon_E\, \delta\tilde{\phi}\, \partial_y\delta\tilde{n},
\] 
which adds about 40 Joules every second to the system. This term plays  the role of the electron cross-field drift velocity along with the gradient of the fluctuating density in the horizontal direction in  driving  the two-stream instability, where the secondary electric field and density-gradient are anti-parallel to each other causing the growth of the unstable modes.\\
\hfill\\
Due to collisions of electrons with the background, 10 Joules are dissipated every second in the diffusion of the fluctuating electric field, 
\[
D_{\phi} =  \int\! d^2x\,  n_eT_e\rho_e^2\nu_{en}\,  \delta\tilde{\phi}\, \nabla^2\delta\tilde{\phi};
\]
however,  the rest of the energy injected into the system is absorbed into  the plasma internal energy via coupling between the dynamics of the electrons and ions.\\
\hfill\\
The energy that is transferred to the ion dynamical  equation through,
\[
C_{\phi\chi} =  \int\! d^2x\,  n_eT_e\, \delta\tilde{\phi}\nabla^2\delta\tilde{\chi}
\]
dissipates via the ion collisional and viscosity terms through \[
D^{(1)}_{\chi} =- \int\! d^2x\, \frac{4}{3}\frac{n_iT_i}{\nu_{in}} \, \nabla^2\delta\tilde{\chi} \nabla^2\delta\tilde{\chi},
\quad \mathrm{and}\quad 
D^{(2)}_{\chi} =  \int\! d^2x\,  n_im_i\nu_{in}\, \delta\tilde{\chi}\nabla^2\delta\tilde{\chi} 
\]
respectively. It was found that the dissipation in the collisional part is almost double that due to  viscosity,  which explains the importance of the ion collisions with the background compared to the ion viscosity in stabilizing the unstable modes and saturating the evolving fields.\\
\hfill\\
To conclude, the electron dynamics injects energy into the system and the major source of energy comes from the cross-field drift when it exceeds the ion-acoustic speed. The injection of the energy through the density-gradient is small for the current simulation conditions compared to the other source of energy. The collisions  of the ions and electrons with  neutrals provide a large sink for the available energy in the dynamical system. In addition, the ion viscosity dissipates one-third of the transferred energy to the ion nonlinear dynamics,  which is also a stabilizing factor in the linear regime. The coupling between the dynamics of the ions and electrons takes  place through the diffusion of the species.

\section{Small-Scale Structures and Energy Cascading}\label{section:EnergyCascades}
\begin{figure}
 \centering
 \begin{tabular}{c}
 \includegraphics[width=1.0\textwidth,height=0.8\textwidth]{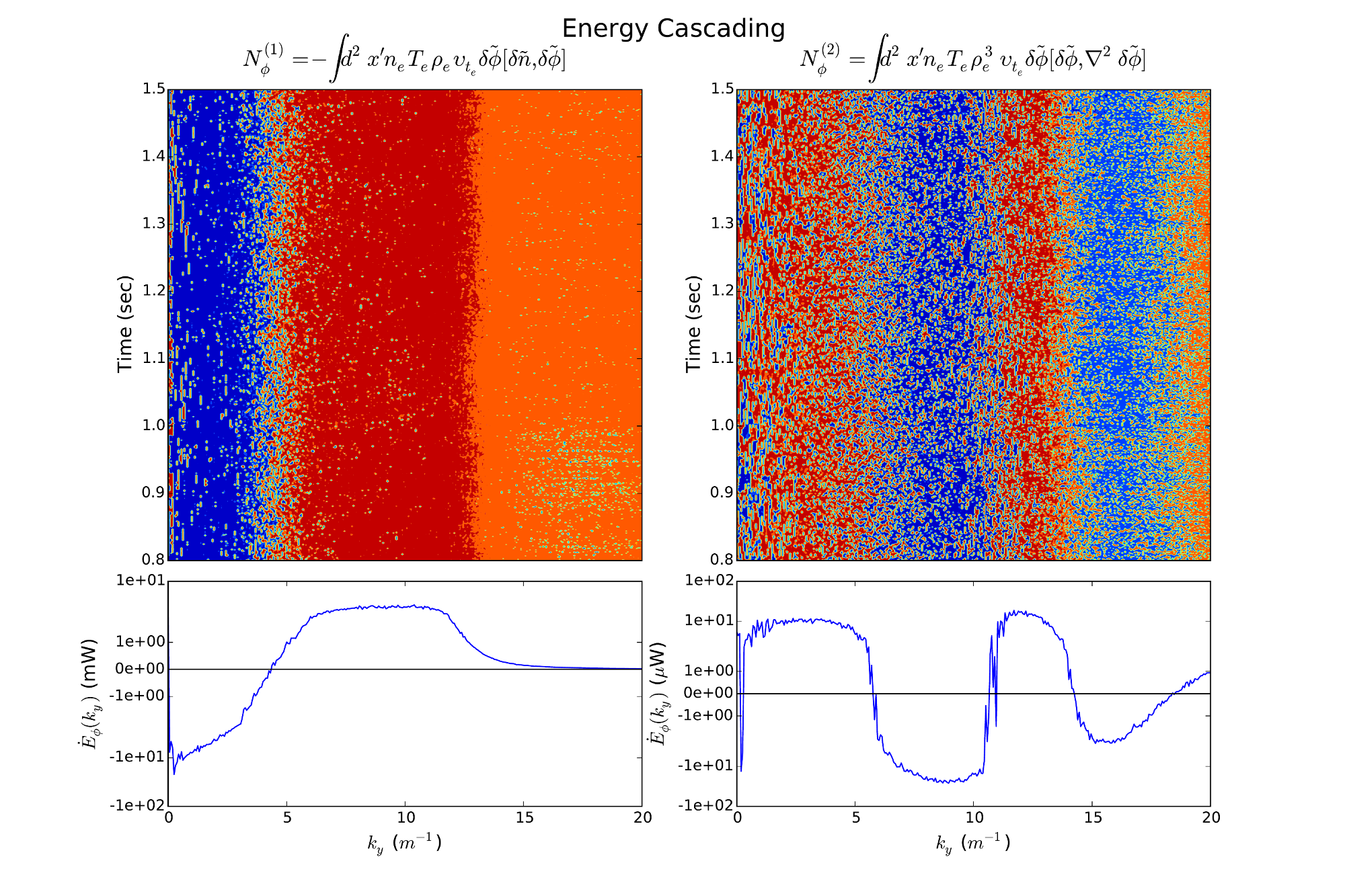}\\
 \end{tabular}
 \caption{\small Forward and dual energy cascades over the saturation state of the simulation for a case of $L_n$ = 6 km and $\upsilon_E$ = 400 m/s. The top panels show the rate of energy transfer over the saturation time as a function of the horizontal ($k_y$) (left) and vertical ($k_z$) (right) wavenumbers. The lower panels show the average rate of energy transfer over the entire saturation regime as a function of $k_y$ and $k_z$. The left-lower panel shows the role of $N_{\phi}^{(1)}$ in transferring the energy from the large structures to the small ones, however the $N_{\phi}^{(2)}$ is responsible of transferring the energy in forward and reverse directions over different regions of wavenumbers but with smaller effect compared to $N_{\phi}^{(1)}$.} 
 \label{fig:CascadeCompare}
\end{figure}

\begin{figure}
 \centering
 \begin{tabular}{c}
 \includegraphics[width=1.0\textwidth,height=0.8\textwidth]{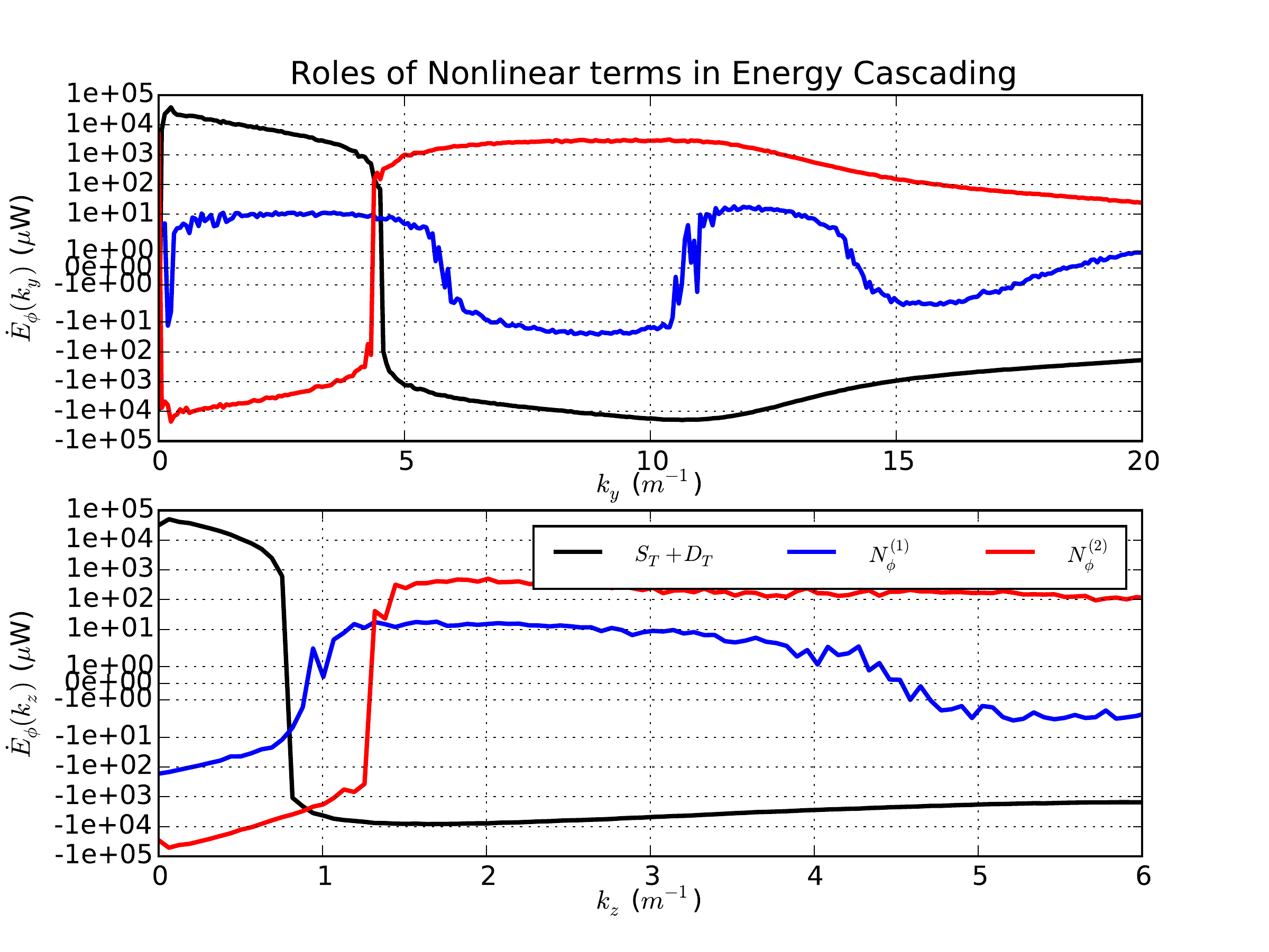}\\
 \end{tabular}
 \caption{\small A comparison between the rate of energy transfer in the two nonlinear terms and the sum of the source and sink terms of energies over the saturation state of the simulation for a case of $L_n$ = 6 km and $\upsilon_E$ = 400 m/s. This emphasizes on the responsibility of the nonlinear energy transitional terms ($N_{\phi}^{(1)}$ and $N_{\phi}^{(2)}$) in generating the small structures in the dynamical system which can not be generated linearly due to the absence of energy sources at these wavenumbers.} 
 \label{fig:CascadeVsSDT}
\end{figure}

\begin{figure}
 \centering
 \begin{tabular}{c}
 \includegraphics[width=1.0\textwidth,height=0.8\textwidth]{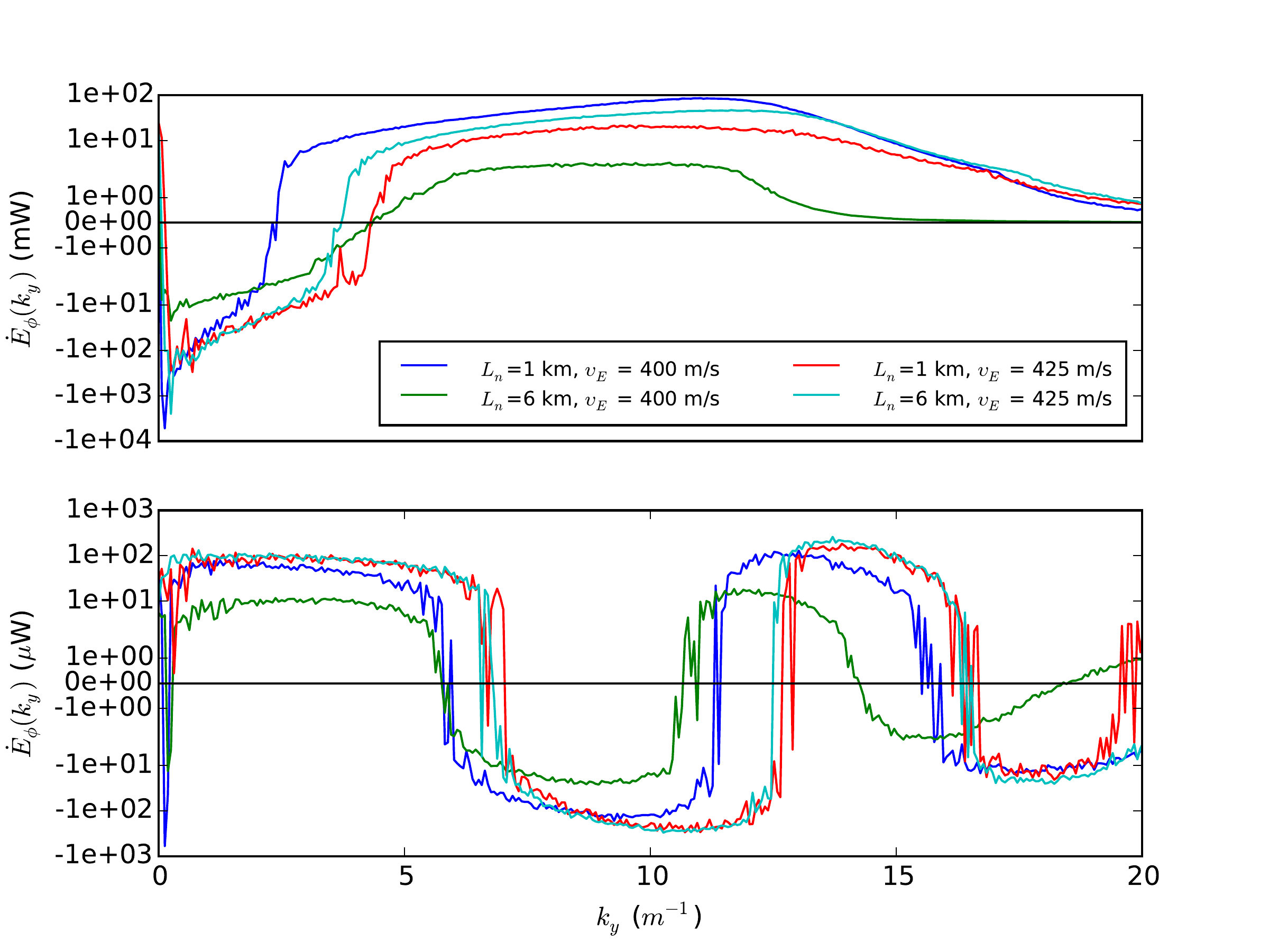}
 \end{tabular}
 \caption{\small A comparison between the forward and reverse energy cascades over the saturation state of the simulation for different cases of $L_n$ = 1 and 6 km and $\upsilon_E$ = 400 and 425 m/s.} 
 \label{fig:AllCascades}
\end{figure}

In section \ref{section:EnergyConservation} we identified three kinds of energy:   sources that inject  energy into the system, sinks that suck energy out of the system, and  (nonlinear) transfer terms that  redistribute energy between structures of different scales in the system.  The transfer terms are  responsible for the generation of plasma structures that cannot be generated linearly.
In equation (\ref{eq:EnergyContributions}) we identified two nonlinear terms, $N_{\phi}$, whose sum over k-space should vanish at each time step, yet they are responsible for transferring  energy between the stable/unstable modes.  These two transfer terms come from the  equation for  the perturbed electric potential, which elucidates the role of the electron dynamics in cascading  energy between the modes at different wavenumbers and, thereby,  generating multi-scale plasma structures.\\
\hfill\\
The first transfer term 
\begin{equation}
N^{(1)}_{\phi} =   \int\! d^2x\, \frac{en_o}{B}\, \delta\tilde{\phi}[\delta\tilde{\phi},\delta\tilde{n}]
\label{nphi}
\end{equation}
is shown  during the saturation phase of the simulation in the left column of Figure~\ref{fig:CascadeCompare} for $L_n$ = 6 km and $\upsilon_E$ = 400 m/s. The color contour plot in the top-panel shows this term integrated in the vertical direction as a function of time and the horizontal wavenumber ($k_y$), where we can see a steady transfer of energy between the unstable modes over the saturation phase without much variation in time. The lower-panel shows the time-average of the energy content of $N^{(1)}_{\phi}$ of \eqref{nphi} as a function of $k_y$. From these panels we can see a forward energy cascade, where the rate of energy change is negative in the region of small-wavenumber  $k_y\le$ 5.0 $m^{-1}$ (red) and positive in the region above $k_y\ge$ 5.0 $m^{-1}$ (blue).\\
\hfill\\
This indicates that the Poisson bracket $[\delta\tilde{\phi}, \delta\tilde{n}]$ transfers  energy from  long structures of small wavenumber to the small structures of large wavenumber. The energy is transferred in successive cascading processes to smaller structures until it ends at structures of a meter or less with a peak in the range of $k_y$ = 9.0 - 11.0 $m^{-1}$. Above $k_y$ = 15 $m^{-1}$, the energy content is almost zero. These results are in agreement with the radar observations that suggest sub-meter structures with very small energy content scatter the radio-frequency signals, while  all structures above half a meter are expected to be seen in the radar echoes when the proper frequency is used.\\
\hfill\\
The other nonlinear term that is responsible for energy cascading, as shown in the right column of 
Figure~\ref{fig:CascadeCompare}. However, the effect of
\begin{equation}
N^{(2)}_{\phi} =  \int\! d^2x\, \frac{en_o}{B} \, \delta\tilde{\phi}[\delta\tilde{\phi},\nabla^2\delta\tilde{\phi}]
\end{equation}
is much smaller than $N^{(1)}_{\phi}$ due to the small amount of cascading energy  (note the \textit{milliwatts} and \textit{microwatts} units on the $y$-axis in the bottom-left and bottom-right panels in 
Figure~\ref{fig:CascadeCompare}, respectively). The Poisson bracket   $[\delta\tilde{\phi},\nabla^2\delta\tilde{\phi}]$ shows two overlapped regions of forward and inverse energy cascading. The first region again at $k_y$ = 0.0 - 5.0 $m^{-1}$, where the energy is transferred via  a forward cascade from a very narrow region of long-scale structures around the size of the simulation box at $k_y\sim$ 0.1 $m^{-1}$ to the smaller structures of order of 1 - 10 meters.  In the second region at $k_y$ = 1 - 15 $m^{-1}$,  we can see dual energy cascading mechanisms. In the right half of the second region ($k_y\le4\pi$) the energy is inversely cascaded from the 0.75 - 1.00 meter structures to the longer-structures of order 1 - 10 meters. This underscores the fact that the Farley-Buneman instability itself can generate structures of long-wavelength as proposed by Kudeki \textit{et al.}~[1987] in the top-part of the E-region that has a negative density scale-length. But this possibility here is limited by the smallness of the amount of energy inversely cascaded to  large-scale plasma structures. On the other half of the second region we can see a forward energy cascade to irregularities of scale-length less than  half a meter. Upon checking the peaks and the spectral spread in the two halves  of the second region,  we can see that the effect of the forward cascading in the second region is stronger than that of the reverse cascading that has a smaller peak and gets energy from the two regions in independent energy cascading processes.\\
\hfill\\
To ensure that the forward and inverse energy cascades are the major factors in generating the small and large plasma irregularities, respectively, in the equatorial electrojet, we made a comparison in Figure~\ref{fig:CascadeVsSDT} between the energy cascading terms and the free energy (the sum of energy sources and dissipations) available in the system dynamics in the horizontal ($k_y$) and vertical ($k_z$) directions. It can be seen in Figure~\ref{fig:CascadeVsSDT} that the sources and sinks largely cancel, leaving the nonlinear transfer terms to redistribute the energy into a state where a balance can occur. This emphasizes the role of the nonlinear terms in generating the unstable plasma waves of small wavelength in the equatorial electrojet when there is no source of free energy is available in the domain of small structures to be responsible for generating them.\\
\hfill\\
A similar comparison between the role of  $N^{(1)}_{\phi}$  and $N^{(2)}_{\phi}$ in cascading the energy between the unstable modes of different scales in the vertical direction is seen in the bottom panel of Figure~\ref{fig:CascadeVsSDT}. Both nonlinear terms are responsible for generating small structures through the forward energy cascading mechanism (red and blue solid lines),  where there are no linearly unstable modes available that can explain the generation of these structures. These forward energy cascades transfer the energy content in the plasma waves of long structures to structures of wavelength 1 - 5 meters. These are the effective wavelength seen in the backscattering spectral echoes of the radar observations. This is further evidence for the two-step theory of  Sudan \textit{et al.} [1973]; i.e.\   energy cascading  gives rise to the nonlinear generation of plasma structures of 3 meter length   observed in the equatorial electrojet, which cannot be explained by  linear theory.\\
\hfill\\
It can also be noticed in Figure~\ref{fig:CascadeVsSDT} that there is a positive rate of energy transfer around $k_y$ = 4 $m^{-1}$ in the top panel and a negative rate of energy transfer around $k_z$ = 1 $m^{-1}$ in the lower panel. Although this imbalance in the rate of energy transfer is six order of magnitude less than the average rate of energy transfer in the system, it can cause an increase of energy over time. However, this imbalance in energy transfer between the unstable active modes in the dynamical system is due to the small time-window we have for calculating the energy due to the large cost of computation. But,  a larger time-window will show a large-scale wave of the perturbed electron density (in the saturation regime) with the small-scale fluctuations  superimposed on the top,  which may give rise to a balance in energy transfer over a properly selected longer period of time. The superimposing of the small fluctuations that are caused by the Farley-Buneman instability on the top of large oscillations caused by the gradient-drift instability is in good agreement with the rocket observations by Pfaff \textit{et al.} [1987b].\\
\hfill\\
An interesting comparison between simulation results for  forward and reverse energy cascading mechanisms associated with  the unstable modes as they enter the nonlinear regime  is shown in Figure~\ref{fig:AllCascades},  for two density scale-lengths ($L_n$ = 1 and 6 km) and two electron cross-field drifts ($\upsilon_E$ = 400 and 425 m/s). The first interesting point is the similarity of the cascading  in Figure~\ref{fig:AllCascades} for different magnitudes of the density scale-length  at the same $\bm{\delta{E}\times{B}}$ drift value of $\upsilon_E$=425 m/s, a similarity that is observed in both the upper and lower panels of this figure. While the difference between the rate of energy transfer of the evolving modes in the Poisson  bracket  
$[\delta\tilde{\phi}, \delta\tilde{n}]$ in $N^{(1)}_{\phi}$ at $\upsilon_E$ = 425 m/s is smaller than the difference between the corresponding profiles at $\upsilon_E$ = 400 m/s,  that difference is negligible in the Poisson  bracket $[\delta\tilde{\phi}, \nabla^2\delta\tilde{\phi}]$ in $N^{(2)}_{\phi}$. It is also surprising to find that  the rate of forward energy cascade in the top panel of Figure~\ref{fig:AllCascades} is small in the case of the  sharp density-gradient (i.e.\  small density scale-length) when $\upsilon_E$ = 425 m/s,  in contrast  to the cases where  $\upsilon_E$ = 400 m/s. This comparison shows that the density-gradient scale-length manifests itself in transferring  energy from large-scale structures to small-scale ones when the cross-field drift is not too much greater than  the ion-acoustic speed where $L_n$ does not have a significant effect on the energy cascading.\\
\hfill\\
Thus, the polarization drift of the electrons helps in understanding the generation of structures of order smaller than a meter and in understanding the inverse energy cascade that allows the generation of long-wavelength structures in the absence of the gradient-drift instability conditions in the ionospheric  background. These simulation results substantiate the Sudan \textit{et al.}~[1973] two-step theory for  energy cascading to the small-scale structures, irregularities in the horizontal and vertical directions, of the equatorial electrojet. 

\section{Summary}\label{section:SummaryConclusions}
In this paper we have analyzed and studied properties of the unified fluid model for equatorial electrojet phenomena of Hassan \textit{et al.}~[2015].  Simulations with parameters set to various ionospheric background conditions revealed properties of the gradient-drift and Farley-Buneman instabilities.   The effect of the density-gradient scale-length on the evolution of the  instabilities was examined in both the linear and nonlinear regimes. It was found that the sharper the density-gradient (i.e.\  the shorter the scale-length $L_n$) the larger the growth-rate in the linear regime for the gradient-drift and Farley-Buneman instabilities. It was seen  that the density-gradient scale-length influences  unstable modes of all sizes in the electrojet. In contrast, changes in the cross-field $\bm{{E}\times{B}}$ drift did not have any noticeable effect on  plasma structures of long-wavelength,  but did  have a  strong effect on  small-scale plasma structures.\\
\hfill\\
A parameter scan for determining the  effects of  $L_n$ and $\upsilon_E$ on the linear growth-rate showed five different regions based on the magnitude of the ion-acoustic speed: two upper regions when the cross-field drift exceeds the ion-acoustic speed and three when the cross-field drift is below this value. Type-II plasma instability dominates in the three regions where the  cross-field speed is smaller than the ion-acoustic speed and the magnitude of the density-gradient scale-length controls the  dynamics. However, Type-I plasma instability dominates the regions where the  cross-field drift is much larger than the ion-acoustic speed, where  the density-gradient scale-length does not show a large effect. In the transition region,  which is characterized by cross-field drifts close to the ion-acoustic speed, both the $\upsilon_E$ and $L_n$ plays important roles in driving both types of instabilities.\\
\hfill\\
In the nonlinear simulation results, the cross-field drift of 425 m/s injects more energy in the system and the unstable modes grow faster and reach higher values at the saturation phase of the simulation compared to the results with only 400 m/s cross-field drift for the same values of density-gradient scale-length. It is also found that the density-gradient scale-length has a negligible effect on nonlinear simulation results in the case of 425 m/s drift over all phases of simulation, however, the sharper density-gradient causes a sharp growing of the unstable modes in the linear phase of the simulation at 400 m/s drift.\\
\hfill\\
To better understand the physics of the plasma instabilities,  we studied the total energy budget of the unified fluid model, composed of the  three components: the kinetic energy of electrons and ions and the internal energy of both species. We observed that the internal energy and the ion kinetic energy are linked together and contain about 98\% of the system energy,  while the electron kinetic energy comprises less than 2\% of the total energy. The reason for this link between the density and the ion velocity potential  was traced to the fact that their dynamical  equations are perfectly coupled together.\\
\hfill\\
We also categorized the rate of change of energy into sources, sinks, transfer, and coupling terms, where the sources of the energy entered  the system from the boundaries through  gradients of the background electron density and electric potential, while energy is dissipated by  the viscosity of the electrons and ions as they collide with the background neutrals.  As an important verification of the code, we determine that the system conserves energy and has very good balance between energy sources and sinks.   \\
\hfill\\
In order to understand the generation of structures of various scales (both horizontal and vertical) in the electrojet, we examined the nonlinear transfer of energy.  Forward and reverse dual energy cascading was found in the system in the horizontal direction. The Poisson bracket  $[\delta\tilde{\phi},\delta\tilde{n}]$ was observed to be  responsible for a large forward cascade from  structures of large wavelength (small wavenumbers) to generate plasma waves of short wavelength. The other Poisson  bracket   $[\delta\tilde{\phi}, \nabla^2\delta\tilde{\phi}]$ was seen to be  three-orders of magnitude less effective than the first. However, it was seen to produce concurrent forward and inverse energy cascading  between structures of different scales. This Poisson  bracket is responsible for the second step of forward  cascading of  energy to   further smaller structures in the equatorial electrojet and inversely cascading  energy that  generates  plasma waves of large sizes. In the vertical direction,  forward energy cascading dominates the dynamics in both nonlinearities  and was seen to be  responsible for the generation of the 2-5 meter structures, which  cannot be generated by  linear dynamics.  Our  study of the forward and inverse energy cascading was seen to substantiate  the two-step energy cascading theory proposed by Sudan [1973].\\
\hfill\\
We presented new achievements in the coupled instabilities in a dynamic system that conserves energy and forms a non-canonical Hamiltonian system. Also, we verified the two-step energy cascading mechanism in the equatorial electrojet region and studied closely the effect of the density-gradient scale-lengths and cross-field drifts on the evolution and saturation of the turbulent plasmas. The concentration of the energy available in the system in the small structures agrees with the radar observations that show that the spectrum of the backscattered echoes is dominated by these small wave structures.


%
%
%
%
%
%
%
\begin{acknowledgments}
Hassan, E. would like to thank Ioannis Keramidas Charidakos for valuable feedback and fruitful discussions. Hatch, D.R. and Morrison,  P.J. were supported by the US Department of Energy Contract No. DE-FG02-04ER54742.
\end{acknowledgments}
\end{article}



\clearpage

\clearpage

\clearpage

\clearpage

\clearpage

\clearpage

\clearpage


\clearpage

\clearpage

\clearpage

\clearpage

\clearpage

%
%

%
%
%
%
%
%
%



\begin{thebibliography}{}
Balsley, Ben B.  ``Some characteristics of non-two-stream irregularities in the equatorial electrojet."  Low-Frequency Waves and Irregularities in the Ionosphere. Springer Netherlands, 1969. 152-172.\\
Bowles, K. L., B. B. Balsley, and Robert Cohen. ``Field-aligned E-region irregularities identified with acoustic plasma waves." J. Geophys. Res., 68.9 (1963): 2485-2501.\\
Farley, D. T., and B. B. Balsley (1973), ``Instabilities in the equatorial electrojet," J. Geophys. Res., 78,227.\\
Farley, D. T. ``The equatorial E-region and its plasma instabilities: a tutorial." Ann. Geophys 27.4 (2009): 1509-1520.\\
Fejer, B.G., Farley, D.T., Balsley, B.B., and Woodman, R.F. (1975). ``Vertical structure of the VHF backscattering region in the equatorial electrojet and the gradient drift instability." J. Geophys. Res., 80, 1313.\\
Hassan, E., W. Horton, A. I. Smolyakov, D. R. Hatch, and S. K. Litt. ``Multiscale equatorial electrojet turbulence: Baseline 2‐D model." J. Geophys. Res.: Space Physics 120, no. 2 (2015): 1460-1477.\\
Hassan, E., I. Keramidas Charidakos, D.R. Hatch, P.J. Morrison, W. Horton, ``Plasma Turbulence in the Equatorial Electrojet: 2-D Fluid Model Hamiltonian" under preparation (2016). \\
Kelley, M.C. (2009). The Earth's Ionosphere, Plasma Physics and Electrodynamics, 2nd edition, Amsterdam ; Boston : Academic Press, 2009.\\
Kudeki, E., Fejer, B.G., Farley, D.T., and Hanuise, C. (1987). ``The CONDOR Equatorial Electrojet Campaign: Radar results," J. Geophys. Res., 92, 13,561.\\
Kudeki, E., and Farley, D.T. (1989). ``Aspect Sensitivity of Equatorial Electrojet Irregularities and Theoretical Implications," J. Geophys. Res., 94, A1, 426-434.\\
Lu F., Farley, D.T., and W.E. Swartz (2008). ``Spread in aspect angles of equatorial E region irregularities," J. Geophys. Res., 113, A11309, doi:10.1029/2008JA013018.\\
Morrison, Philip J. ``Hamiltonian description of the ideal fluid." Rev. Mod.  Phys. 70.2 (1998): 467.\\
Pfaff, R. F., M. C. Kelley, E. Kudeki, B. G. Fejer, and K. Baker (1987a), ``Electric field and plasma density measurements in the strongly driven daytime equatorial electrojet, 1, The unstable layer and gradient drift waves", J. Geophys. Res., 92, 13578.\\
Pfaff, R.F., Kelley, M.C., Kudeki, E., Fejer, B.G., and Baker, K.D. (1987b). ``Electric field and plasma density measurements in the strongly driven daytime equatorial electrojet. 2 Two-stream waves," J. Geophys. Res., 92, 13,597.\\
Pope, Stephen B. Turbulent flows. Cambridge university press, 2000.\\
Schmidt, M. J., and S. P. Gary (1973), ``Density gradients and the Farley-Buneman instability," J. Geophys. Res., 78, 8261.\\
Sudan, R. N., J. Akinrimisi, and D. T. Farley. ``Generation of small‐scale irregularities in the equatorial electrojet." J. Geophys. Res., 78.1 (1973): 240-248.\\
Sudan, R. N. ``Nonlinear theory of type I irregularities in the equatorial electrojet." Geophys. res. letters 10.10 (1983a): 983-986.\\
Sudan, R. N. ``Unified theory of type I and type II irregularities in the equatorial electrojet." J. Geophys. Res.: Space Physics (1978–2012) 88.A6 (1983b): 4853-4860.\\

%
%
%
%

\end{thebibliography}
\end{document}